\documentclass[useAMS,usegraphicx,usenatbib]{mn2e}
\usepackage{amssymb}

\title[Different morphology and kinematics in HH~666]{HH~666: Different kinematics from H$\alpha$ and [Fe~II] emission provide a missing link between jets and outflows}

\author[Reiter et al.]{Megan Reiter$^{1}$\thanks{E-mail: mreiter@as.arizona.edu (MR)}, Nathan Smith$^{1}$, Megan M. Kiminki$^{1}$, and John Bally$^{2}$ \\
$^{1}$Steward Observatory, University of Arizona, Tucson, 933 N. Cherry Ave, Tucson, AZ 85721, USA \\
$^{2}$Center for Astrophysics and Space Astronomy, University of Colorado, 
    389 UCB, Boulder, CO 80309, USA} 

\begin{document}

\date{Accepted 2014 Nov ??. Received 2014 Nov ??; in original form 2014 November ??}

\pagerange{\pageref{firstpage}--\pageref{lastpage}} \pubyear{2014}

\maketitle

\label{firstpage}


\begin{abstract}
HH~666 is an externally irradiated protostellar outflow in the Carina nebula for which we present new near-IR [Fe~{\sc ii}] spectra obtained with the FIRE spectrograph at Magellan Observatory. 
Earlier H$\alpha$ and near-IR [Fe~{\sc ii}] imaging revealed that the two emission lines trace substantially different morphologies in the inner $\sim 40$\arcsec\ of the outflow. 
H$\alpha$ traces a broad cocoon that surrounds the collimated [Fe~{\sc ii}] jet that extends throughout the parent dust pillar. 
New spectra show that this discrepancy extends to their kinematics. 
Near-IR [Fe~{\sc ii}] emission traces steady, fast velocities of $\pm 200$ km s$^{-1}$ from the eastern and western limbs of the jet. 
We compare this to a previously published H$\alpha$ spectrum that reveals a Hubble-flow velocity structure near the jet-driving source. 
New, second-epoch \textit{HST}/ACS H$\alpha$ images reveal the \textit{lateral} spreading of the H$\alpha$ outflow lobe away from the jet axis. 
H$\alpha$ proper motions also indicate a sudden increase in the mass-loss rate $\sim 1000$ yr ago, while steady [Fe~{\sc ii}] emission throughout the inner jet suggest that the burst is ongoing. 
An accretion burst sustained for $\sim 1000$ yr is an order of magnitude longer than expected for FU~Orionis outbursts, but represents only a small fraction of the total age of the HH~666 outflow. 
Altogether, available data suggests that [Fe~{\sc ii}] traces the highly collimated protostellar \textit{jet} while H$\alpha$ traces the entrained and irradiated \textit{outflow}. 
HH~666 appears to be a missing link between bare jets seen in H~{\sc ii} regions and entrained molecular outflows seen from embedded protostars in more quiescent regions. 
\end{abstract}



\section{Introduction}\label{s:intro}
Over a wide range of initial masses, stars appear to drive an outflow at some stage in their formation. 
Mounting evidence calls into question whether there is an upper limit on the mass of protostars that may drive collimated jets, despite uncertainty about whether the dominant physics of formation may change between low and high masses \citep[e.g.][]{gar03,car08,guz12,dua13,rei13}. 
These jets will interact with and shape -- and be shaped by -- the circumstellar molecular envelope and larger natal cloud. 
Various jet properties and morphologies, such as asymmetries \citep[e.g.][]{rei98,bal01}, wiggles and gaps \citep[e.g.][]{deG96}, the degree of collimation \citep{mun91}, and decreasing knot velocities have been explained by interaction with different kinds of environments.

Interaction between the protostellar jet and ambient gas in the star forming cloud has been proposed as a mechanism to create molecular outflows. 
\citet{rag93c} offer a simple analytical model wherein molecular outflows are the product of environmental gas being entrained by a protostellar jet. 
While some molecular outflows are well-described by jet-entrainment \citep[see, e.g.][]{zin98,arc05,arc06}, others show a conical morphology and broader range of velocities in the outflowing gas that have been interpreted as a wide-angle wind launched from the protostellar disc \citep[e.g.][]{nag97,kla13,sal14}. 
A full evolutionary scenario may involve a combination of these two, with different components dominating at different times \citep[e.g.][]{sha06,zap14}. 
In either case, the outflow will clear material along the jet axis, creating a polar cavity as it travels through the protostellar envelope and into the surrounding molecular cloud. 
Clearing of the protostellar envelope may explain the apparent widening of molecular outflows associated with more evolved sources \citep{arc06}.

\begin{table*}
\caption{Observations\label{t:obs}}
\centering
\begin{tabular}{llllll}
\hline\hline
Instrument & Filter / Position & Date & Int. time & Comment & Ref. \\
\hline
ACS/\textit{HST} & F658N & 2005 Mar 30 & 1000 s & H$\alpha$ $+$ [N~{\sc ii}] & \citet{smi10} \\
\vspace{3pt}
ACS/\textit{HST} & F658N & 2014 Feb 21 & 1000 s & H$\alpha$ $+$ [N~{\sc ii}] & this work \\
WFC3-UVIS/\textit{HST} & F656N & 2009 Jul 24 & 7920 s & H$\alpha$ & \citet{rei13} \\
WFC3-UVIS/\textit{HST} & F673N & 2009 Jul 24 & 9600 s & [S~{\sc ii}] $\lambda6717$ $+$ $\lambda6731$  & \citet{rei13} \\
WFC3-IR/\textit{HST} & F126N & 2009 Jul 25 & 600 s & [Fe~{\sc ii}] $\lambda12567$ & \citet{rei13} \\
\vspace{3pt}
WFC3-IR/\textit{HST} & F164N & 2009 Jul 25 & 700 s & [Fe~{\sc ii}] $\lambda16435$ & \citet{rei13} \\
\vspace{3pt}
EMMI/\textit{NTT} & H$\alpha$ & 2003 Mar 11 & 2400 s & P.A. $=293.5^{\circ}$ & \citet{smi04} \\
FIRE/\textit{Magellan} & HH~666~M & 2013 Feb 22 & 600 s & P.A. $=110^{\circ}$ & this work \\
FIRE/\textit{Magellan} & HH~666~IRS & 2013 Feb 22 & 600 s & P.A. $=110^{\circ}$ & this work \\
FIRE/\textit{Magellan} & HH~666~O & 2013 Feb 22 & 600 s & P.A. $=115^{\circ}$ & this work \\
\hline
\end{tabular}
\label{t:foobar}
\end{table*}
%

In the unobscured environment outside the natal cloud, optical and near-IR emission lines can be used as a tracer of the protostellar \textit{jet} \citep[e.g.][]{sep11}. 
Longer wavelength emission lines from shock-excited molecules, on the other hand, sample embedded parts of the entrained \textit{outflow} propagating within a molecular cloud \citep[e.g.][]{nor04}. 
In this paper, we refer to the \textit{jet} as the fast, highly collimated stream of gas launched near the poles of the protostar and the \textit{outflow} as the slower, wide-angle component of outflowing gas that either originated in the protostellar disc, or in the ambient medium that was entrained by shocks in the passing jet. 
Outflows tend to be observed in more embedded regions \citep[e.g.][]{bel08,beu08,for09,fue09,tak12}. 
The underlying collimated jets are primarily atomic and remain invisible unless excited by shocks or external irradiation. 
For both excitation mechanisms, bare jets emerging from more evolved, unembedded sources will emit at visual and UV wavelengths \citep[e.g.][]{rei98,bal01,mcg04}. 
Observing both the jet and the outflow in a single source is difficult, but understanding how the jet interacts with the surrounding cloud is important for constraining local momentum feedback. 
Partially embedded jets offer one way to bridge these two regimes. 
For example, a protostar residing near the edge of a cloud or in a small, isolated globule may drive a jet into the relatively diffuse environment outside the cloud. 
At the same time, the counter-jet plows deeper into the molecular cloud where its presence must be inferred from the shape of the outflow lobe it creates. 
Jets and outflow lobes have been seen to coexist in several protostellar sources (e.g. HH~111: \citealt{nag97,lef07}; HH~46/47: \citealt{nor04,arc13}; and HH~300: \citealt{rei00,arc01a}).

In the Carina nebula, HH~666 emerges from a translucent pillar into a giant H~{\sc ii} region created by more than $\sim 65$ O-type stars \citep[see][]{smi06a}. 
\citet{smi04} discovered the outflow in ground-based narrowband optical images and identified several large knots that extend more than 4 pc on the sky (features HH~666~D, A, E, M, O, N, I, and C). 
The mass-loss rates \citep{smi10,rei13} and outflow velocities \citep{rei14} measured in HH~666 are among the highest of the jets in Carina, making it one of the most powerful jets known \citep[see, e.g. Figure~14 in][]{ell13}. 
Harsh UV radiation from the many massive stars in Carina permeates the region, illuminating the dust pillar from which HH~666 emerges and irradiating the body of the jet itself. 
External irradiation lights up neutral material in the jet that is not excited in shocks, revealing portions of the jet that would otherwise remain invisible. 
This allows the physical properties of the jet to be derived using the diagnostics of photoionized gas \citep[e.g.][]{bal06,smi10}. 
However, massive star feedback illuminates indiscriminately, irradiating both the jet and the surrounding environment, which can complicate the interpretation of the observed emission. 
Non-uniform density structures in a photoionized jet can further complicate the interpretation. 
Fortunately, multi-wavelength observations allow us to disentangle the jet from its complex environment.

\begin{figure*}
\centering
\includegraphics[trim=-4mm -3mm 0mm 0mm,angle=0,scale=0.850]{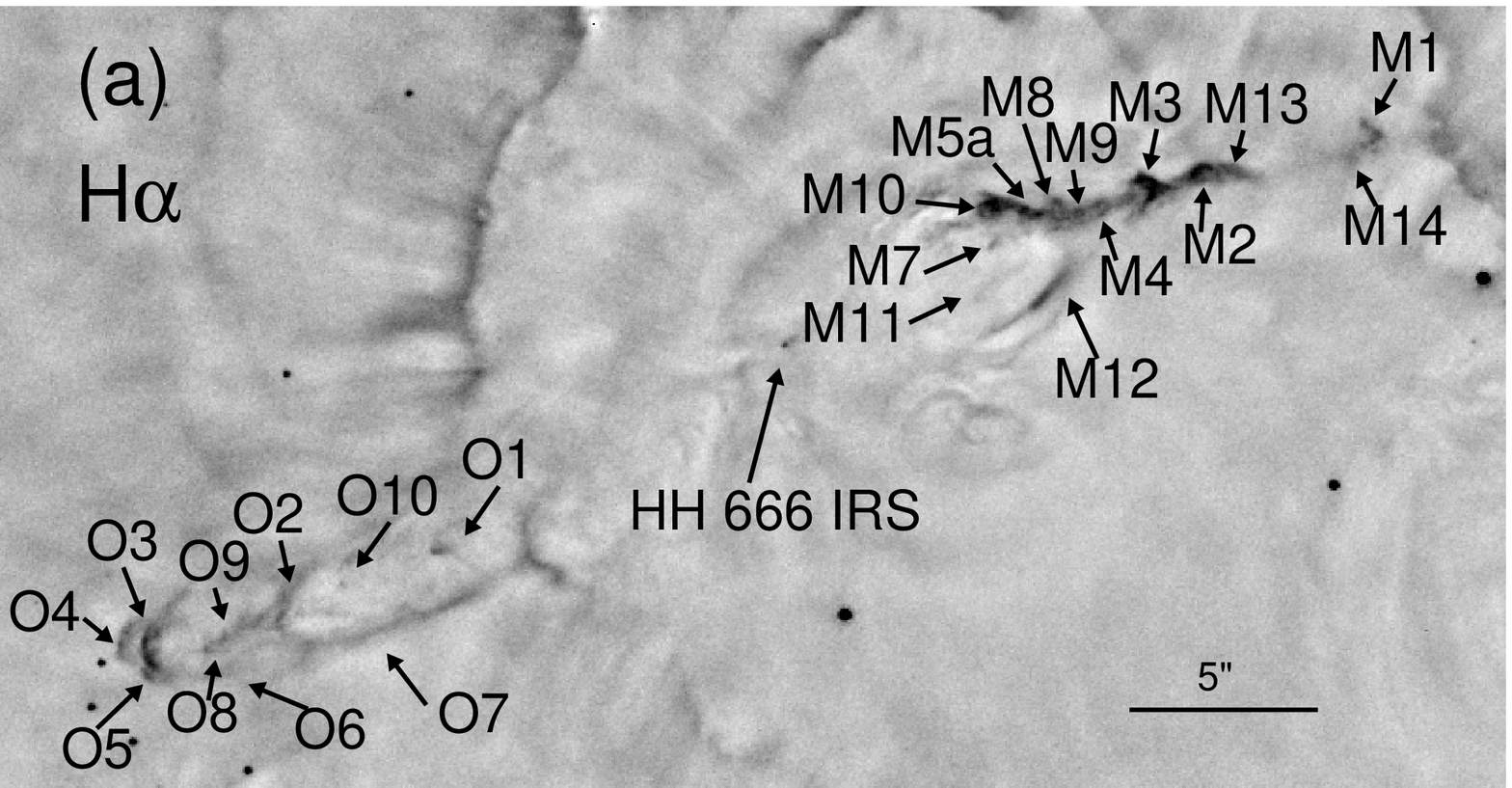}
$\begin{array}{cc}
\includegraphics[trim=15mm 0mm 0mm 0mm,angle=0,scale=0.380]{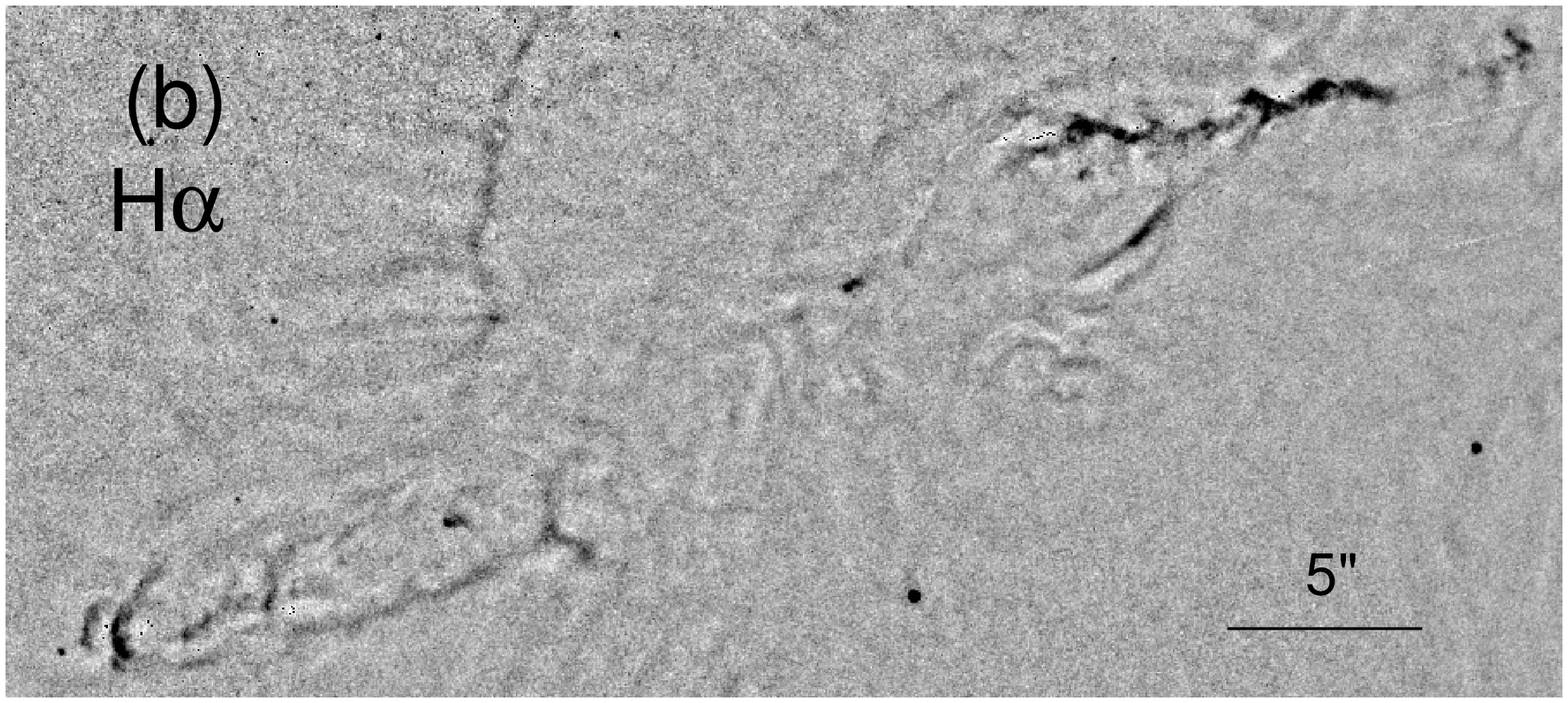} &
\includegraphics[trim=45mm 0mm 10mm 0mm,angle=0,scale=0.380]{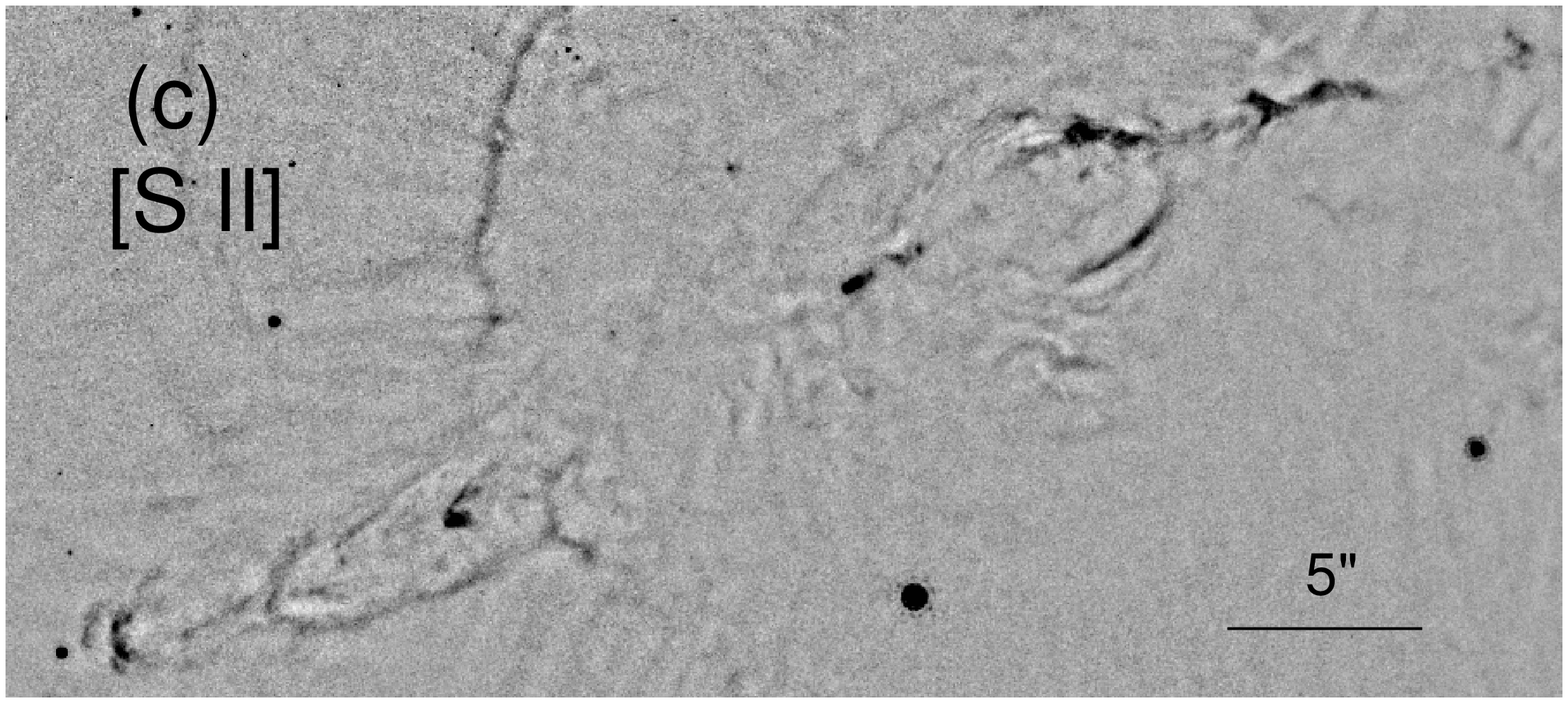} \\ 
\includegraphics[trim=0mm 0mm 0mm 0mm,angle=0,scale=0.35]{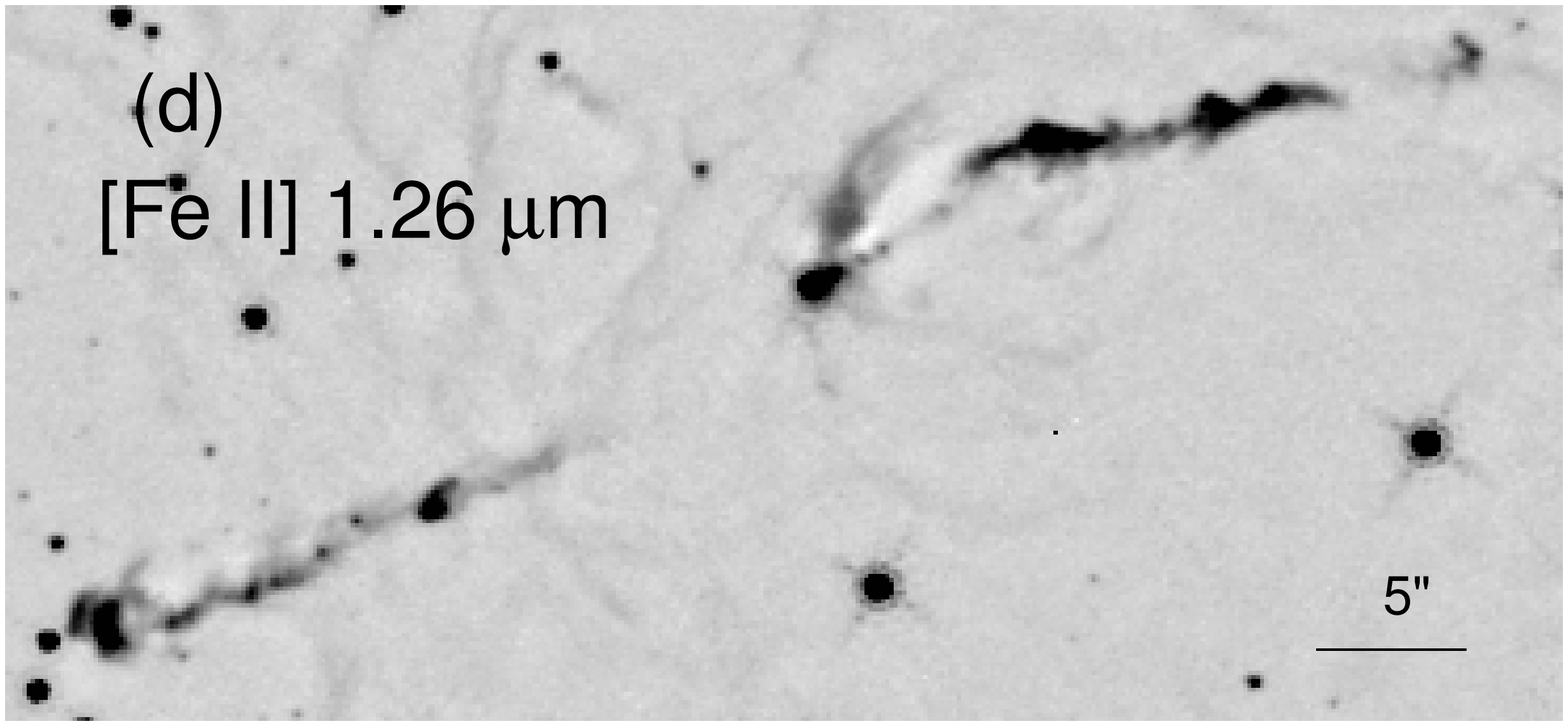} &
\includegraphics[trim=20mm 0mm 0mm 0mm,angle=0,scale=0.35]{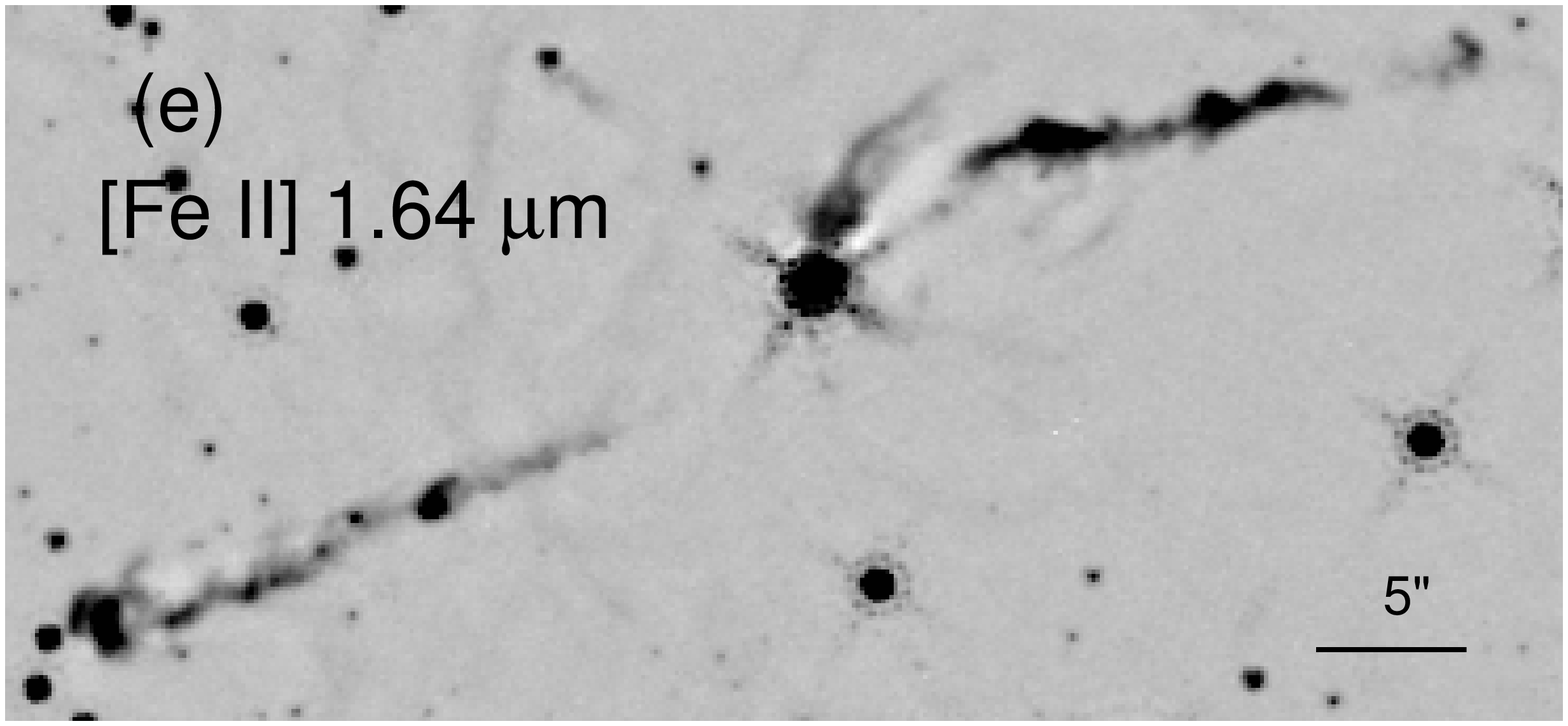} \\ 
\end{array}$
\includegraphics[trim=80mm 0mm 80mm 12mm,angle=90,scale=0.70]{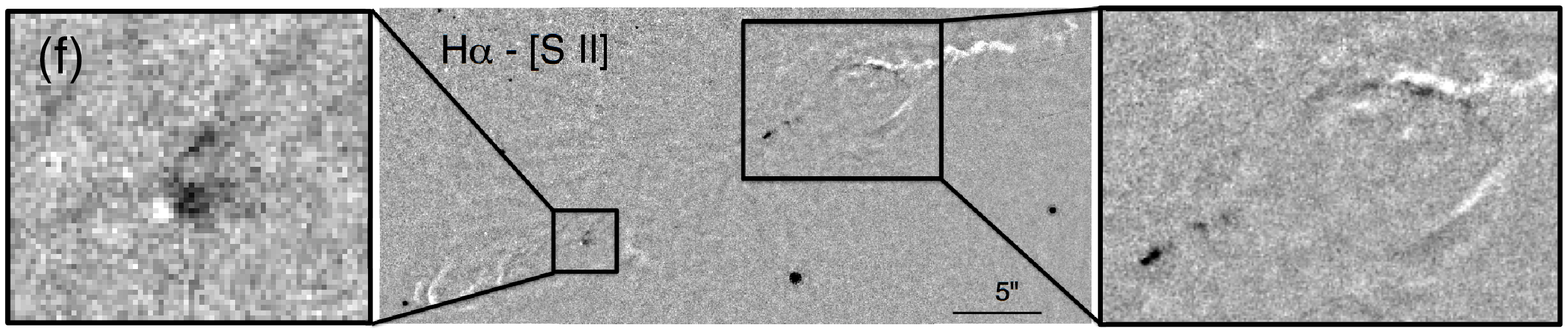}
\caption{
\textit{Top:} New \textit{HST}/ACS H$\alpha$ image of HH~666 with labels and arrows pointing out features discussed in the text and used to measure proper motions (a). 
\textit{Middle:} \textit{HST}/WFC3-UVIS H$\alpha$ (b) and [S~{\sc ii}] (c) images from \citet{rei13} with an unsharp mask applied to improve the contrast between faint jet features and bright emission from the surface of the dust pillar. 
Narrowband, near-IR [Fe~{\sc ii}] images with an unsharp mask applied to improve contrast are shown in the row below (panels d and e). 
\textit{Bottom:} An H$\alpha$-[S~{\sc ii}] difference image (f) showing the offset between the two emission lines (excess H$\alpha$ emission appears white, [S~{\sc ii}] emission is black). Small images on either side show a zoomed-in view of the two places in the innermost jet where the two emission lines are significantly offset. 
}\label{fig:hst_ims} 
\end{figure*}

\citet{smi04} and \citet{rei13} showed that near-IR [Fe~{\sc ii}] emission from HH~666 traces the jet inside the $\sim 40$\arcsec\ wide pillar, connecting the large-scale jet seen in H$\alpha$ to the IR-bright protostar that drives it (HH~666~IRS). 
High angular resolution near-IR narrowband [Fe~{\sc ii}] images obtained with Wide-Field Camera 3 (WFC3-IR) on the \textit{Hubble Space Telescope (HST)} reveal additional morphological complexities \citep[see Figure~\ref{fig:hst_ims} and][]{rei13}. 
Near the driving source, [Fe~{\sc ii}] emission traces portions of the jet that are not seen in H$\alpha$. 
In the western limb of the inner jet, designated HH~666~M, [Fe~{\sc ii}] emission bisects a dark cavity to the northwest of HH~666~IRS, clearly connecting the jet to its driving source. 
[Fe~{\sc ii}] emission from HH~666~O, the eastern limb of the inner jet, traces a collimated jet body that appears to be surrounded by a broad cocoon of H$\alpha$ emission. 
Curiously, \citet{smi04} see a Hubble-like velocity structure in H$\alpha$ and [S~{\sc ii}] spectra of the inner parts of HH~666 (features HH~666~M~and~O). 
Only two other protostellar jets have shown Hubble-like velocity structures in optical spectra (HH~83, \citealt{rei89}; and HH~900, \citealt{rei15}).

In this paper, we present new ground-based near-IR [Fe~{\sc ii}] spectroscopy and a second epoch of \textit{HST}/ACS H$\alpha$ images. 
When combined with extant data, the \textit{HST} images allow us to compare the kinematics of these two morphologically distinct components. 
Together, these new data offer a self-consistent explanation for the curious morphological and kinematic differences between the H$\alpha$ and [Fe~{\sc ii}] emission from HH~666.


\section{Observations}\label{s:obs}
\subsection{Near-IR spectroscopy}\label{ss:spectroscopy}
Near-IR spectra of HH~666 were obtained with the Folded-Port Infrared Echellette spectrograph \citep[FIRE,][]{sim13} on the Baade 6.5-m Magellan telescope on 22 February 2013. 
Conditions were photometric with thin clouds and the seeing was $\sim 0\farcs6$.  
FIRE captures medium resolution spectra from $0.8-2.5$ \micron\ simultaneously. 
The echelle slit is $7$\arcsec\ long and our data were obtained with a slit width of $0\farcs6$. 
This corresponds to a velocity resolution of 50 km s$^{-1}$. Uncertainties in the measured radial velocities are shown in Figure~\ref{fig:spectra}. 
We used five slit positions to target the brightest [Fe~{\sc ii}] emission in the inner $\sim40$\arcsec\ of the HH~666 jet and its driving source. 
Slit positions are shown in Figure~\ref{fig:spectra}. 

We obtained separate frames offset $\sim 5$\arcsec\ from the jet axis for the four pointings along HH~666~M and HH~666~O to subtract sky lines and emission from the H~{\sc ii} region. 
For the central slit position that contains the driving source (see Figure~\ref{fig:spectra}), we employ an ABBA nod putting the driving source at opposite ends of the slit and remove sky emission by subtracting complementary nods. 
The total integration time for each slit position is 600 s. 
The data were reduced using the {\sc firehose IDL} pipeline. 
This data reduction package provides flat-fielding, sky-subtraction, and wavelength calibration of the 2D FIRE spectra. 
Wavelength calibration was done using a ThAr lamp. 
We extract the [Fe~II] position-velocity diagrams from the reduced, calibrated 2D spectra. 
The reported velocities are in the heliocentric velocity frame.

\subsection{New H$\alpha$ images}\label{ss:newacs}
We present new H$\alpha$ images obtained with \textit{HST}/ACS on 21 February 2014. These images duplicate the observational setup used by \citet{smi10} to observe HH~666 on 30 March 2005, providing two epochs with a time baseline of 8.9 yr (see Table~1). 
Both epochs were observed using the F658N filter (which transmits both H$\alpha$ and [N~{\sc ii}] $\lambda6583$) with a total exposure time of 1000 s. A three-position linear offset pattern was used to fill ACS chip gaps. 
See \citet{smi10} for a more complete description of the observational setup. 

By replicating the observational setup of the first-epoch images, we reduce systematic uncertainties in the image registration and measurement of individual blob proper motions. 
Our image registration procedure is similar to the method described in \citet{and08a,and08b}, \citet{and10}, and \citet{soh12}.  
This method uses PSF photometry of pipeline-calibrated images that have been bias-subtracted, flat-fielded, and CTE-corrected, but have not been resampled to correct for geometric distortion (denoted with the file extension {\tt flc}). 
These images are better suited for high-accuracy PSF fitting than drizzled images (denoted by the file extension {\tt drc}), and are therefore preferable to measure precise relative proper motions from \emph{HST} data.  

We construct an initial reference frame from the {\tt drc} images using the astrometric information in their headers.  Basic centroid positions of bright, relatively-isolated stars in the {\tt drc} images are compiled into a reference frame that has a 50 mas pixel$^{-1}$ scale and is aligned with the $y$ axis pointing north, like that of \citet{and08a}.  
We then perform PSF photometry on each {\tt flc} exposure using the program {\tt img2xym\_WFC.09x10} \citep{and06}, which employs a spatially-variable library of PSFs.  
Stellar positions are corrected afterward for geometric distortion using the distortion correction from \citet{and05}.  
Next, six-parameter linear transformations are computed from each distortion-corrected {\tt flc} frame to the initial reference frame via the positions of the stars found in both.  For improved accuracy, we recalculate the transformation after replacing the initial reference-frame positions of the stars with their average PSF-based positions transformed from the {\tt flc} frames.

Finally, we construct a stacked image for each epoch by resampling the {\tt flc} images into the reference frame, allowing us to measure the motions of extended sources.  The details of the stacking algorithm can be found in \citet{and08a}. 
As both epochs are transformed to the same reference frame, object positions are directly comparable.  
The image alignment precision is primarily limited by the internal accuracy of the reference frame and distortion correction \citep[$\sim 0.015$ pixels, see][]{and06,soh12}, which over our time baseline corresponds to $0.084$ mas yr$^{-1}$ or 0.92 km s$^{-1}$ at the distance of Carina \citep[2.3 kpc,][]{smi06}.

To measure proper motions of bright knots in the inner jet, we use our implementation of the modified cross-correlation technique used by \citet[][see also \citealt{mor01}, \citealt{har01}, and \citealt{rei14}]{cur96}. 
First, we correct for local diffuse H~{\sc ii} region emission by subtracting a median-filtered image from each epoch. 
Next, we extract bright jet knots using a box size that is optimized for each feature (see Figure~\ref{fig:pm}). 
To determine the pixel offset between the two epochs, we generate an array that contains the total of the square of the difference between two images for each shift of a small box containing the jet feature relative to a reference image. 
The best match between the two images will produce the smallest value in the array. 

With the identical observational setup and longer time baseline between the two epochs of \textit{HST}/ACS data, we achieve finer velocity resolution than that obtained in our previous study \citep{rei14} using \textit{HST}/WFC3-UVIS images taken 4.32 yr after the first-epoch \textit{HST}/ACS images. 
Using the same camera, filter, and approximate location of the source on the detector (and thus same distortion correction) minimizes systematic uncertainties and allows us to measure proper motions of bright nebular jet features with velocities greater than $\sim 10$ km s$^{-1}$. 
To determine the uncertainty in the pixel offset between the two epochs, we explore the proper motion obtained using different box sizes around each jet feature. We also determine the maximum offset between the two images that still results in good alignment, identified by residuals in a subtraction image on the order of the noise. 
From this, we determine an average proper motion uncertainty of $\sim 10$ km s$^{-1}$, although this depends on the signal-to-noise of the feature (see Table~2).

\subsection{Previously published images and spectroscopy}
For comparison to the new H$\alpha$ image and [Fe~II] spectra, we include \textit{HST}/WFC3 narrowband H$\alpha$, [S~{\sc ii}], and [Fe~{\sc ii}] 1.26 \micron\ and 1.64 \micron\ images. These were discussed previously by \citet{rei13}. 
Full observational details can be found in that paper. 

We also include the H$\alpha$ spectrum of HH~666 from \citet{smi04}. 
Longslit H$\alpha$ observations were obtained over a slit length of $\sim6$\arcmin. 
With the cross-dispersing grism and a $1$\arcsec\ slit width, the H$\alpha$ spectrum has a velocity resolution of $\sim10$ km s$^{-1}$. 
More complete details of the observation and reduction are given by \citet{smi04}.


\section{Results} 

\subsection{Morphology in images}
High angular resolution images from \textit{HST} reveal the complex morphology of HH~666, particularly in the inner $\sim 40$\arcsec\ of the jet. 
H$\alpha$ emission from HH~666~M begins $\sim 10$\arcsec\ to the northwest of the driving source tracing a broad arch around a possible evacuated cavity (see Figure~\ref{fig:hst_ims}) before quickly focusing into a narrow beam. 
On the opposite side of the driving source (to the southeast), a few H$\alpha$-bright knots trace the inner jet. 
Two bright parabolic arcs with extended wings reach back toward the driving source creating a broad cocoon morphology (features O2 and O3 in Figure~\ref{fig:pm}). 
Applying an unsharp mask to the H$\alpha$ image reveals collimated jet emission just inside the terminus of HH~666~O (inside feature O3, see Figures~\ref{fig:hst_ims} and \ref{fig:pm}). 
Comparing H$\alpha$ and [S~{\sc ii}] images from \textit{HST}/WFC3-UVIS demonstrates that the morphology of these two lines is remarkably similar in externally irradiated HH jets \citep{rei13}. 
In purely shock-excited gas, however, the location and relative intensity of the two lines will be offset \citep{har94,hea96}. 
Thus, in the harsh UV environment of the Carina nebula, photoionization dominates the excitation of the jet, leading to significantly brighter H$\alpha$ emission compared to [S~{\sc ii}], with the typical ratio in HH~666 of [S~{\sc ii}]/H$\alpha \sim 0.03$. 

There are only two places in the jet where bright [S~{\sc ii}] emission diverges from the morphology of the H$\alpha$ emission (see Figure~\ref{fig:hst_ims}, bottom panel). 
[S~{\sc ii}] emission extends inside the first arc of H$\alpha$ emission from the jet on either side of HH~666~IRS. 
To the northwest (HH~666~M), [S~{\sc ii}] emission falls just inside the arc of H$\alpha$ emission that appears to trace a cavity, extending $\sim 0\farcs25$ behind the H$\alpha$. 
On the other side of the driving source in HH~666~O, [S~{\sc ii}] is much more pronounced in a difference image, extending nearly 1\arcsec\ behind the H$\alpha$ bright knot that marks the emergence of HH~666~O. 
In a post-shock cooling zone, bright [S~{\sc ii}] emission will trail an H$\alpha$-bright arc, with the specific geometry depending on the density and ionization fraction of the preshock gas, and the velocity of the shock itself \citep{har94,hea96}. 
Such offsets have been observed in high-spatial resolution images of nearby HH jets, and interpreted as spatially resolved post-shock cooling zones \citep[e.g.][]{hea96,rei97,har11}. 
Therefore, interpreting HH~666 as a purely photoionized jet would be an oversimplification.

\begin{figure}
\centering
$\begin{array}{c}
\includegraphics[angle=0,scale=0.375]{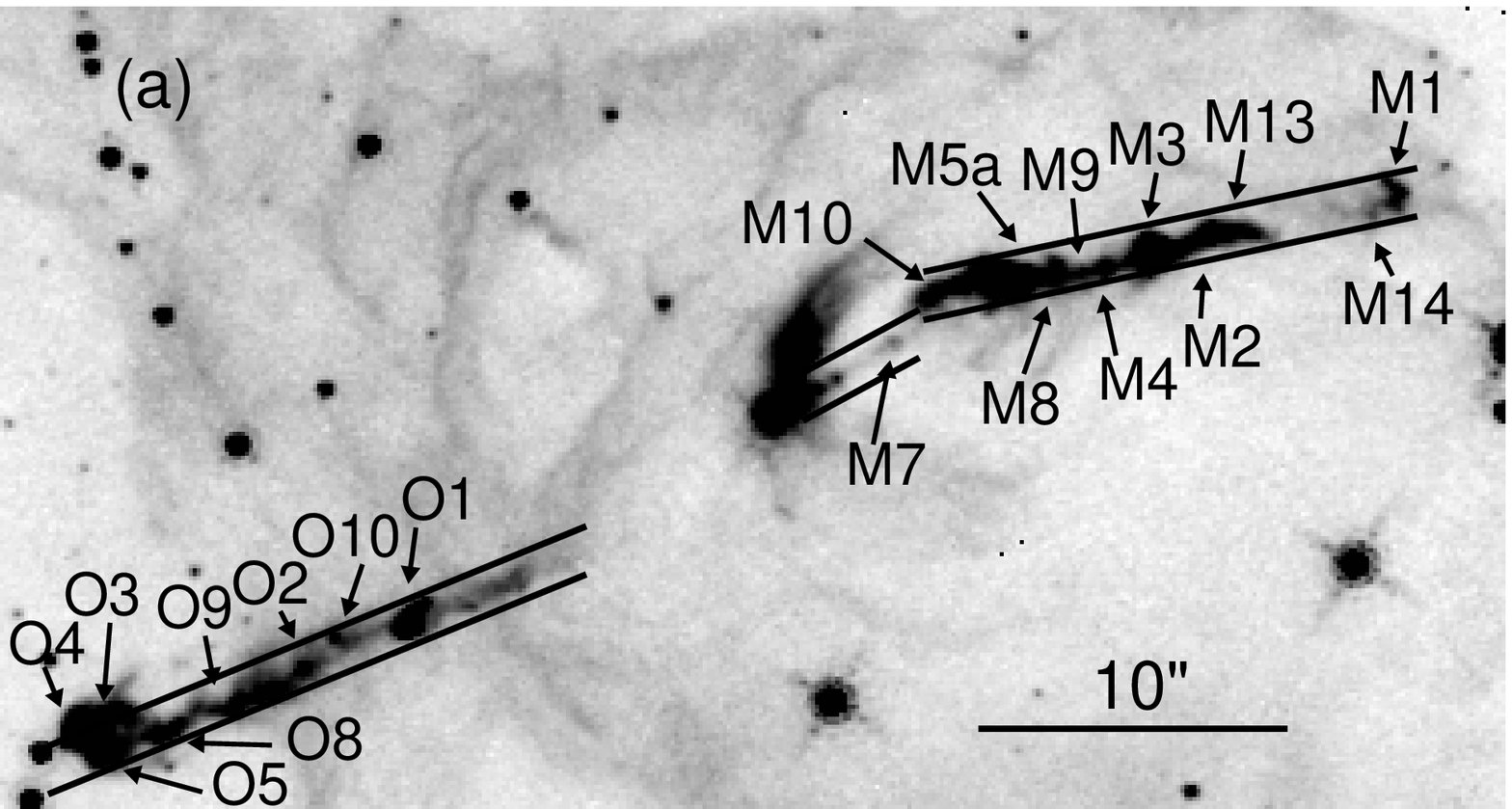}  \\
\includegraphics[trim=0mm 0mm 0mm 15mm,angle=90,scale=0.34]{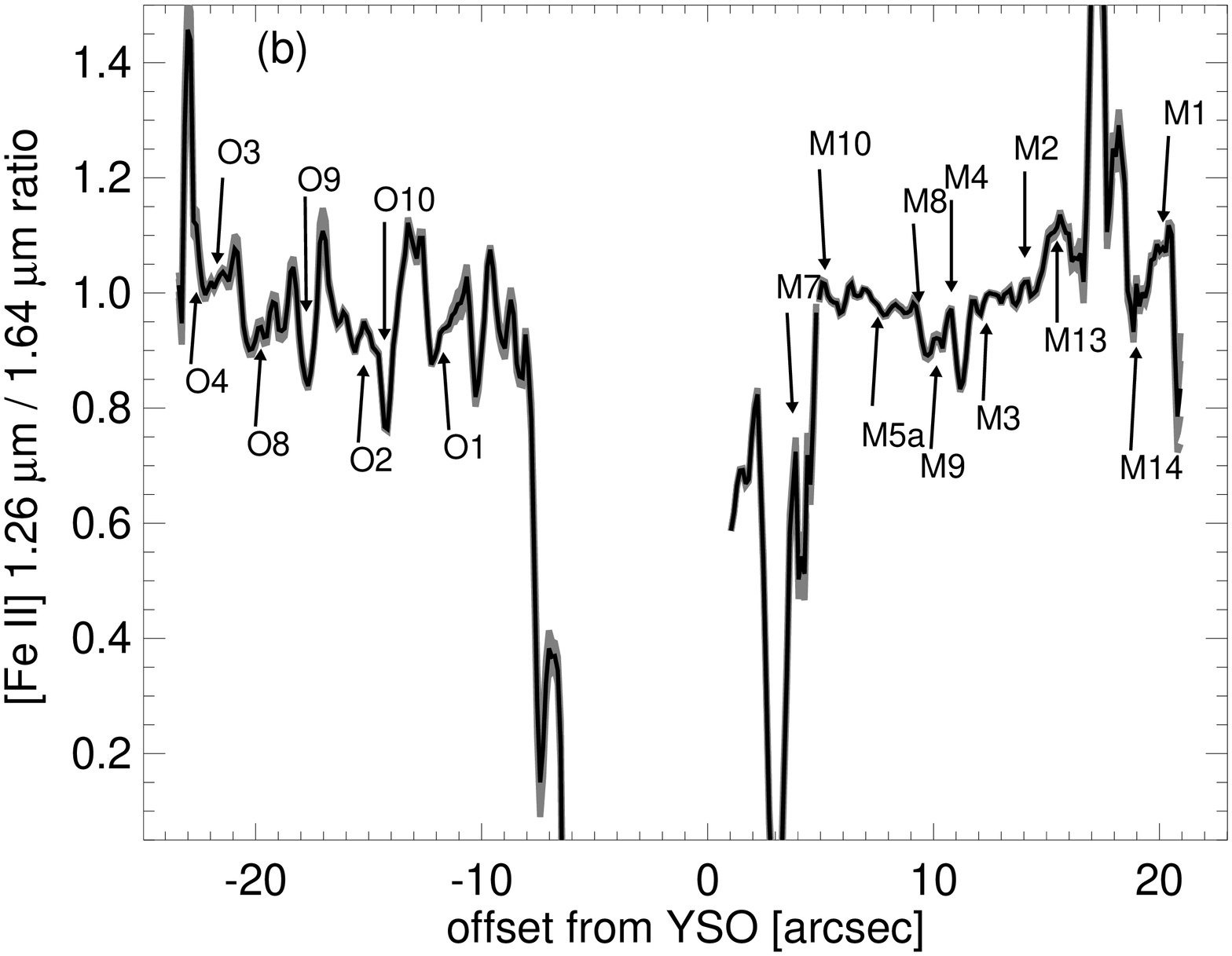}  \\
\includegraphics[trim=0mm 0mm 0mm 15mm,angle=90,scale=0.34]{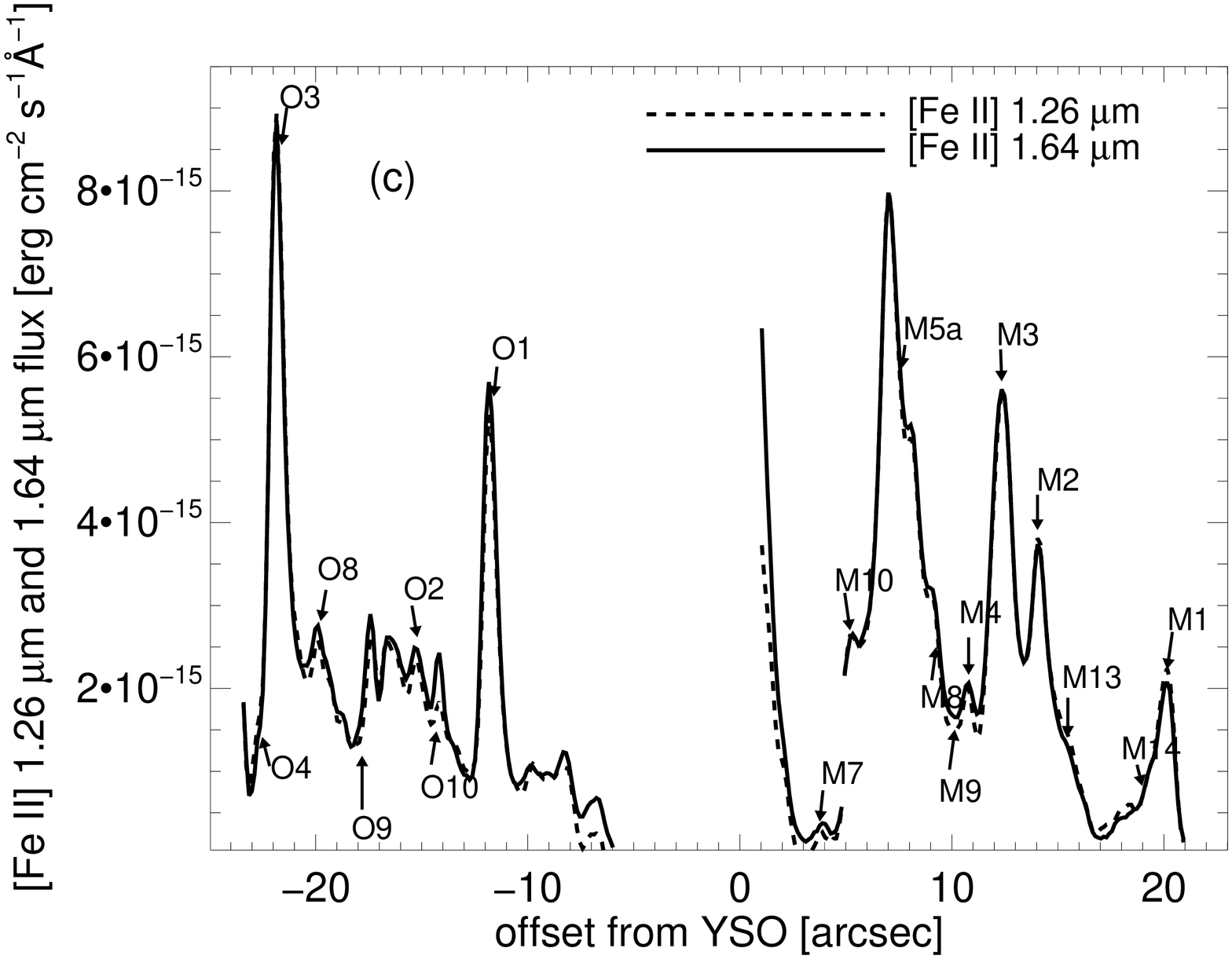}  \\
\end{array}$
\caption{(a) [Fe~{\sc ii}] 1.26 \micron\ image of HH~666 with boxes showing the tracing through the jet to measure the [Fe~{\sc ii}] ratio $\mathcal{R}$ plotted below. 
(b) Plot of the [Fe~{\sc ii}] 1.26 \micron\ / 1.64 \micron\ ratio $\mathcal{R}$ along the inner jet. The shaded area surrounding the ratio line indicates the standard deviation. 
(c) Tracing of the [Fe~{\sc ii}] 1.26 \micron\ and [Fe~{\sc ii}] 1.64 \micron\ flux through the same portions of the inner jet. 
}\label{fig:feii_ratio} 
\end{figure}
Near-IR [Fe~{\sc ii}] emission from HH~666 tells a remarkably different story from that told by the optical emission \citep{rei13}. 
A stream of [Fe~{\sc ii}] emission in HH~666~M bisects the dark cavity seen in H$\alpha$ images, connecting the jet to the driving source. 
Outside the H$\alpha$-bright arc bordering the cavity (features M8, M5a, M10, M12), [Fe~{\sc ii}] emission follows the same morphology as H$\alpha$. 
Morphological discrepancies are even more striking on the opposite side of the jet in HH~666~O. 
[Fe~{\sc ii}] emission does not connect the H$\alpha$ jet all the way to the driving source, instead beginning $\sim 6$\arcsec\ away from the driving source, closer than the $\sim 11$\arcsec\ offset in the H$\alpha$ emission. 
From this point, [Fe~{\sc ii}] emission, although clumpy, extends continuously to the bright shock just outside the pillar edge. 
There is no H$\alpha$ or [S~{\sc ii}] emission from the collimated jet inside the first H$\alpha$ arc in HH~666~O (feature O2 in Figure~\ref{fig:pm}). 
Outside this point, H$\alpha$ and [S~{\sc ii}] emission trace both the cocoon and the jet. 
Optical emission from the outer jet traces the same morphology as the [Fe~{\sc ii}] jet that bisects the larger `sheath.' 
H$\alpha$, [S~{\sc ii}], and [Fe~{\sc ii}] emission are all bright at the tip of HH~666~O just outside the edge of the parent pillar (features O9, O8, O3, O4, O5 in Figure~\ref{fig:pm}).

\subsection{[Fe~{\sc ii}] ratio}
We also use the [Fe~{\sc ii}] images from \citet{rei13} to measure the flux ratio of the two observed lines, $\mathcal{R} = \lambda12567$/$\lambda16435$. 
Both lines originate from the same upper level, so their intrinsic ratio is determined by atomic physics. 
The observed ratio will decrease as the column of absorbing material increases, so $\mathcal{R}$ provides a measure of the reddening. 
\citet{sh06} measured $\mathcal{R} = 1.49$ in P~Cygni while more recent measurements of HH~1 by \citet{gia15} yielded $\mathcal{R} = 1.11 - 1.20$. 
Since we do not have simultaneous off-line continuum images, we estimate local continuum emission from adjacent regions in the [Fe~{\sc ii}] images themselves. 
We plot $\mathcal{R}$ along the length of the inner jet in Figure~\ref{fig:feii_ratio}.

The median flux ratio through the jet is $\mathcal{R} \sim 0.98 \pm 0.15$. 
This corresponds to an A$_V$ of at least $\sim 0.85$ mag assuming $\mathcal{R} = 1.49$ and $\sim 0.25 - 0.41$ mag assuming $\mathcal{R} = 1.11 - 1.20$, respectively. 
Both sides of the jet show an increase in the ratio along the jet moving toward the edge of the pillar. 
The flux ratio is lowest near the driving source, with a median $\mathcal{R} \sim 0.63$, corresponding to A$_V \gtrsim 1.75$ mag for the \citet{sh06} value of $\mathcal{R}$, or A$_V \gtrsim 1.15 - 1.31$ mag using the \citet{gia15} values. 
In contrast, the ratio is closest to the intrinsic value where we do not observe strong [Fe~{\sc ii}] emission from the jet (e.g. outside feature O4 and between features M13 and M14 in Figure~\ref{fig:feii_ratio}).

\subsection{Velocities from spectra}
\begin{figure*}
\centering
$\begin{array}{ccccc}
\includegraphics[trim=15mm 0mm 0mm 0mm,angle=0,scale=0.375]{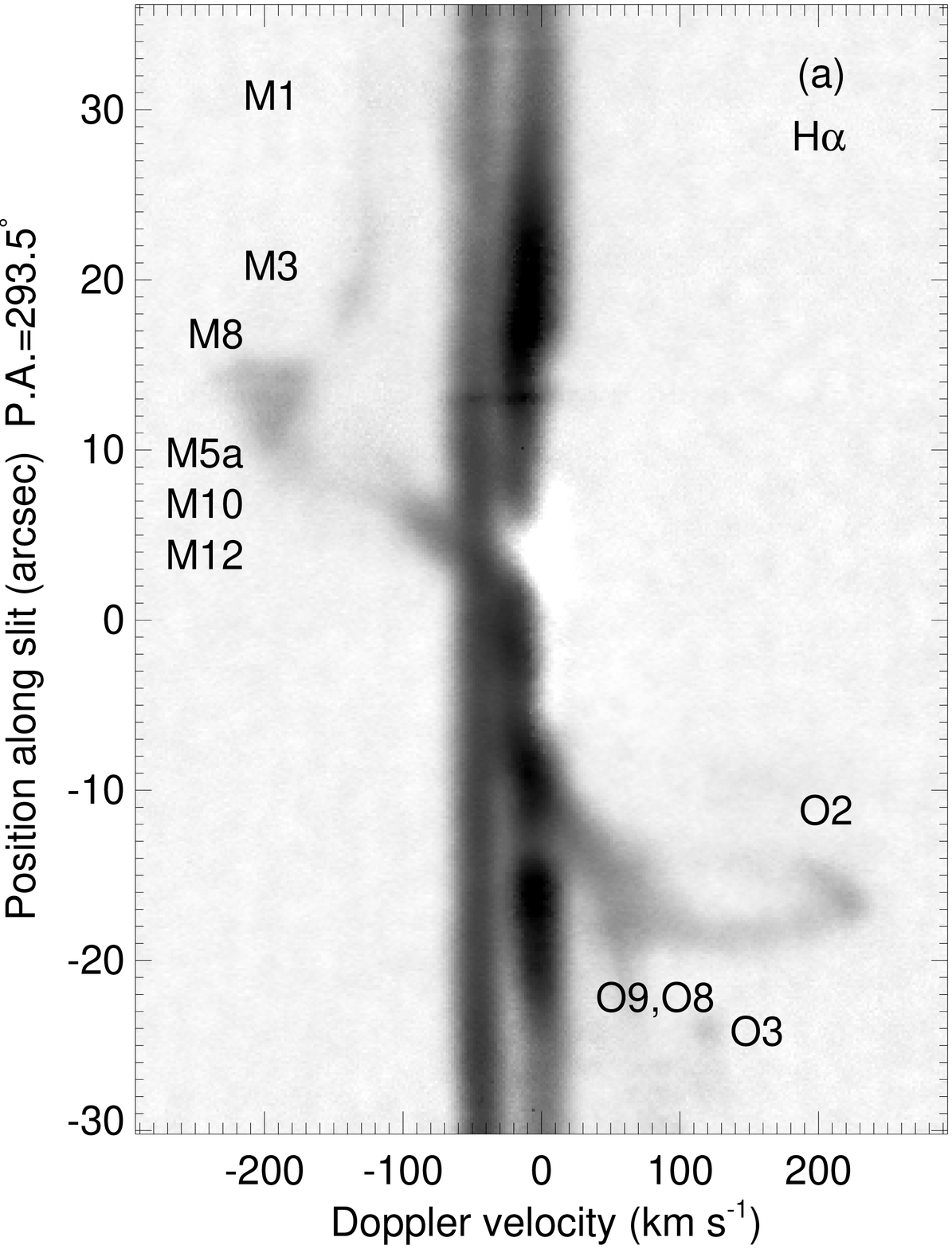} & 
\includegraphics[trim=5mm -7.5mm 8mm 0mm,angle=0,scale=0.385]{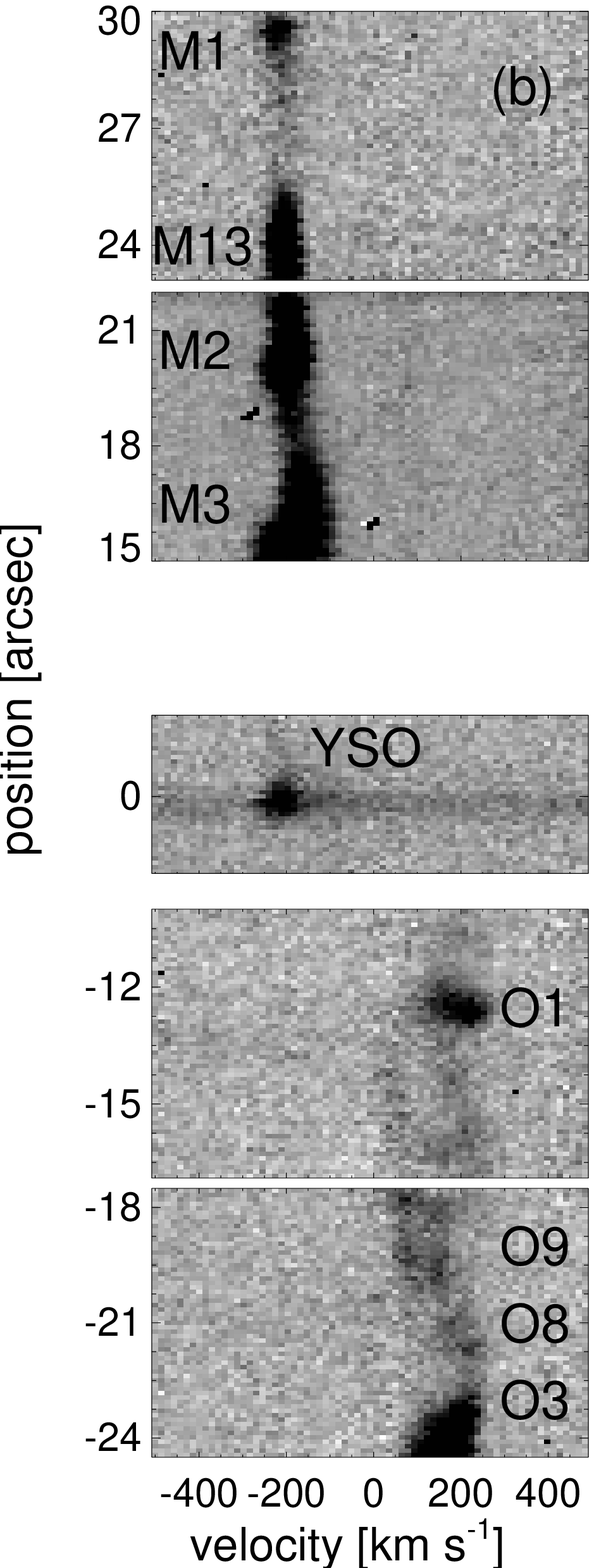} &
\includegraphics[trim=8mm -7.5mm 5mm 0mm,angle=0,scale=0.385]{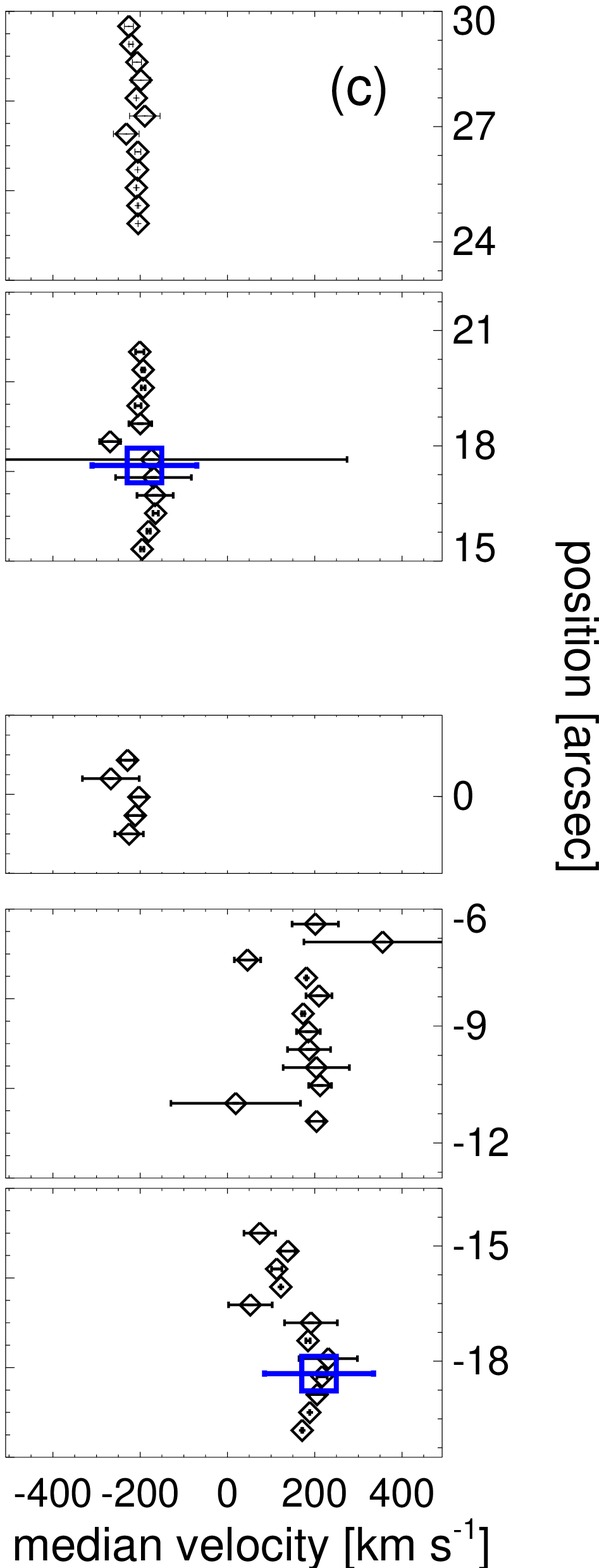} &
\includegraphics[trim=-25mm 0mm 0mm 0mm,angle=90,scale=0.415]{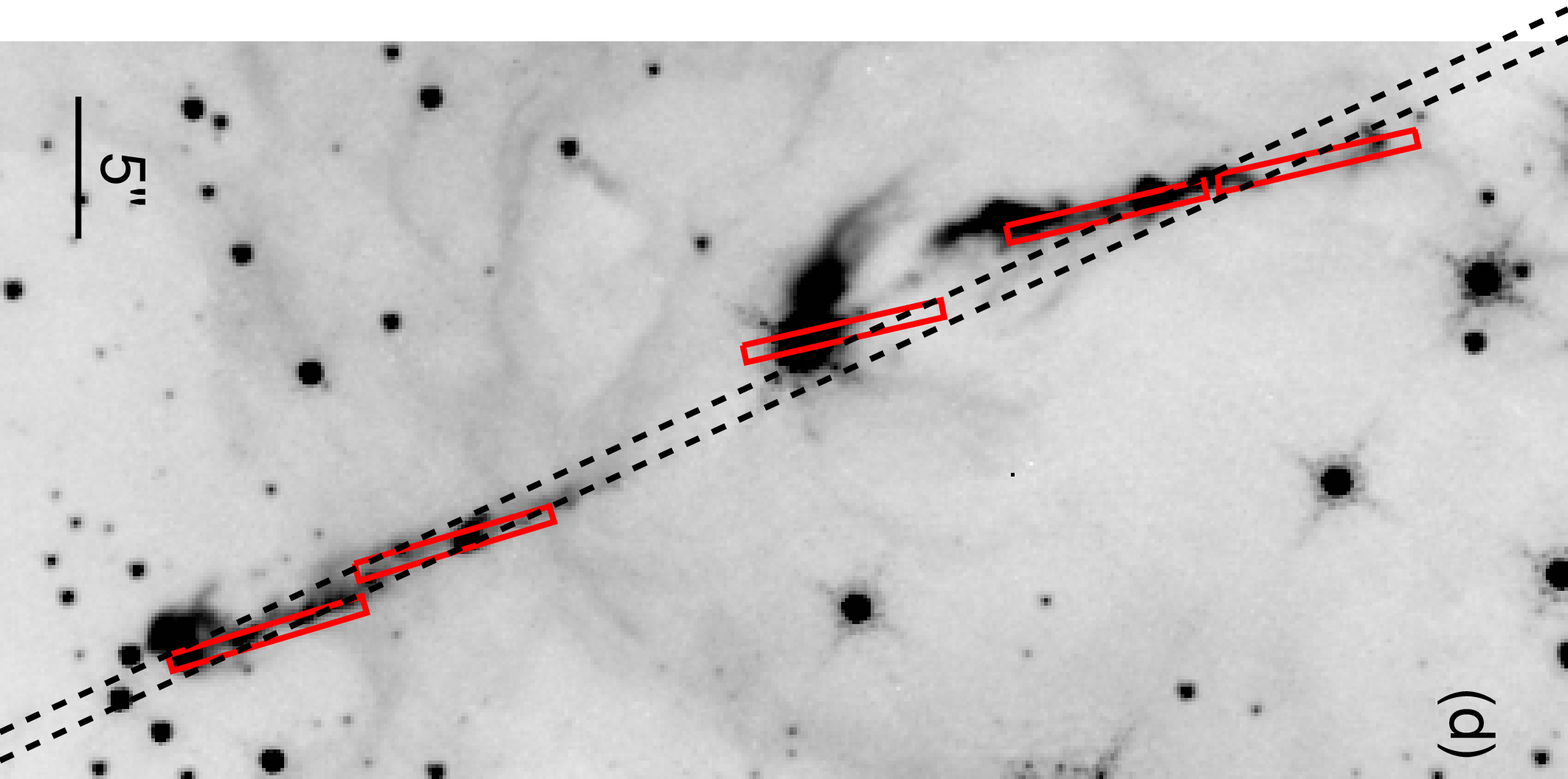} \\
\end{array}$
\caption{
(a) H$\alpha$ position-velocity diagram of HH~666 from \citet{smi04}. 
(b) New, near-IR [Fe~{\sc ii}] 1.26 \micron\ spectra obtained with five separate slit positions shown to the right. Alongside the [Fe~{\sc ii}] position-velocity diagram, we plot the median line velocity over steps of $\sim0\farcs6$ (c). Blue boxes show the H$\alpha$ velocities for HH~666~M and O determined by \citet{smi04}. 
(d) FIRE slit positions shown as red boxes on a [Fe~{\sc ii}] 1.26\micron\ $+$ 1.64 \micron\ image from \textit{HST}/WFC3-IR. The dotted line indicates the slit position used to obtain the longslit H$\alpha$ spectrum. 
}\label{fig:spectra} 
\end{figure*}

In the HH~666 discovery paper, \citet{smi04} published longslit H$\alpha$ and [S~{\sc ii}] spectra that confirmed that bright knots to the east and west of the dust pillar are part of a coherent bipolar outflow. 
Doppler velocities in H$\alpha$ and [S~{\sc ii}] spectra show that the western limb of the jet (including HH~666~M) is blueshifted and the eastern limb (HH~666~O) is redshifted, with velocities from the inner jet reaching $\pm 250$ km s$^{-1}$ \citep{smi04}. 
The rapid increase in velocity away from the driving source on both sides of the jet shows an almost Hubble-like velocity structure. 
In a Hubble flow, velocity is proportional to the distance, $v \propto d$. 
\citet{smi04} derive a ``Hubble constant'' for HH~666 of $H_{HH~666} \simeq 1200 (\mathrm{tan}i)$~km~s$^{-1}$~pc$^{-1}$, where $i$ is the inclination for the fast inner jet. 
A few of the more distant knots in HH~666 also show an increase in velocity with distance from the driving source. 
Both H$\alpha$ and [S~{\sc ii}] emission traces this velocity structure, mirroring the close match of their morphology in images. 

The kinematics traced by [Fe~{\sc ii}] emission are as discrepant as the morphology in images. 
Unlike the H$\alpha$ kinematics of HH~666, the [Fe~{\sc ii}] emission traces steady velocities on either side of the jet. 
An absence of sharp velocity drops that suggests that no strong shocks decelerate the jet. 
Velocities traced by near-IR [Fe~{\sc ii}] are $\pm 200$ km s$^{-1}$ throughout the inner jet (see Figure~\ref{fig:spectra}).

\subsection{New proper motions}
\begin{table*}
\caption{Proper motions in the HH~666 inner jet}
\centering
\begin{tabular}{lrrrrrrr}
\hline\hline
Object & $\delta$x & $\delta$y & v$_T$$^{\mathrm{a}}$ & v$_R$$^{\mathrm{b}}$ & 
velocity$^{\mathrm{c}}$ & $\alpha$ & age$^{\mathrm{d}}$ \\
 & mas & mas & [km s$^{-1}$] & [km s$^{-1}$] & [km s$^{-1}$] & 
[degrees] & yr \\ 
\hline\hline
HH~666~M1 &  -125 (1) &  -34 (1) &    159 (4) &  -104 (6) &   190 (7) &  -33 (2) & 1108 (34) \\
HH~666~M14 &  -122 (3) &  -30 (2) &    154 (5) &  ... &   161 (5) & ... & 1097 (42) \\
HH~666~M13 &  -122 (1) &  -33 (3) &    155 (4) &  ... &   162 (4) & ... & 850 (27) \\ 
HH~666~M2 &  -124 (1) &  -40 (1) &    159 (4) &  ... &   167 (4) & ... & 789 (25) \\
HH~666~M3 &  -113 (0.5) &  -32 (0.2) &    143 (3) &  -131 (7) &   194 (8) &  -42 (1) & 779 (24) \\
HH~666~M4 &  -131 (1) &  -25 (2) &    163 (4) &  ... &   170 (4) & ... & 593 (19) \\
HH~666~M9 &  -126 (2) &  -20 (4) &    156 (4) &  ... &   164 (4) & ... & 568 (19) \\ 
HH~666~M8 &  -123 (1) &  -41 (1) &    159 (4) &  ... &   166 (4) & ... & 537 (17) \\
HH~666~M5a &  -117 (1) &  -31 (2) &    149 (3) &  -193 (6) &   243 (7) &  -52 (1) & 576 (18) \\ 
HH~666~M10 &  -114 (2) &  -37 (1) &    147 (4) &  -193 (6) &   242 (7) &  -53 (1) & 472 (16) \\
HH~666~M7 &  -124 (1) &  -37 (1) &    158 (4) &  ... &   166 (4) & ... & 405 (13) \\ 
HH~666~M11 &  -95 (3) &  -19 (3) &    118 (5) &  ... &   124 (5) & ... & 462 (21) \\ 
HH~666~M12 &  -52 (0.1) &  -7 (0.3) &    64 (1) &  ... &   67 (2) & ... & 1193 (37) \\
HH~666~O1 &   107 (2) &   66 (1) &    154 (4) &  ... &   162 (4) & ... & 741 (24) \\
HH~666~O10 &   123 (3) &   47 (3) &    162 (5) &  ... &   169 (6) & ... & 895 (35) \\
HH~666~O2 &   88 (3) &   63 (6) &    132 (6) &   223 (18) &   259 (19) &   59 (4) & 1249 (66) \\
HH~666~O9 &   117 (3) &   94 (5) &    184 (7) &   57 (8) &   192 (10) &   17 (1) & 989 (41) \\
HH~666~O8 &   134 (2) &   67 (2) &    184 (5) &   57 (8) &   192 (9) &   17 (0.4) & 1017 (34) \\
HH~666~O3 &   108 (1) &   54 (1) &    148 (4) &  ... &   155 (4) & ... & 1363 (44) \\
HH~666~O4 &   91 (1) &   59 (2) &    133 (3) &   121 (4) &   180 (5) &   42 (1) & 1564 (52) \\
HH~666~O5 &   108 (2) &   47 (2) &    144 (4) &  ... &   151 (4) & ... & 1434 (52) \\
HH~666~O6 &   50 (67) &   52 (52) &  88 (73) &  ... &  92 (77) & ... & 2156 (1792) \\ 
HH~666~O7 &   9 (3) &   35 (2) &    45 (3) &  ... &   47 (3) & ... & 3210 (194) \\
\hline
\end{tabular}
\medskip 
\footnotesize 
\begin{tabular}{l}
Uncertainties for each quantity are listed in parenthesis. \\
$^a$ The transverse velocity, assuming a distance of 2.3 kpc, measured over $\Delta t = 8.9$ yr. \\
$^b$ The radial velocity. \\
$^c$ The total space velocity, assuming the average inclination of $42^{\circ}$ where the radial velocity could not be matched. \\
$^d$ Time for the object to reach its current position at the measured velocity, assuming ballistic motion. \\
\end{tabular}
\label{t:pm}
\end{table*}
We compare new ACS H$\alpha$ images obtained in 2014 to the first epoch images obtained $\sim 9$ yrs earlier by \citet{smi10} and measure proper motions of jet features. 
In general, the measured velocities agree with those found by \citet{rei14} within the uncertainties (see Table~2), although the better match of the instrument and filter used in this work permit higher precision than in our previous work.

\begin{figure*}
\centering
$\begin{array}{c}
\includegraphics[angle=0,scale=0.85]{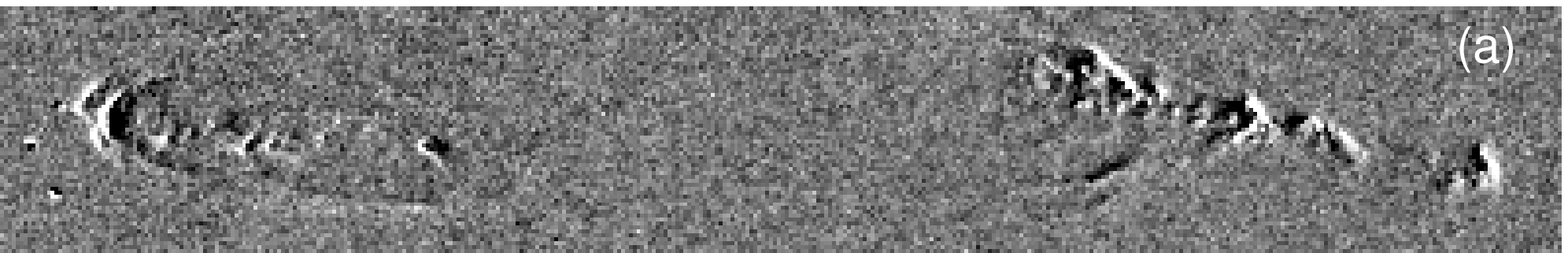}  \\
\begin{array}{cc}
\includegraphics[angle=0,scale=0.335]{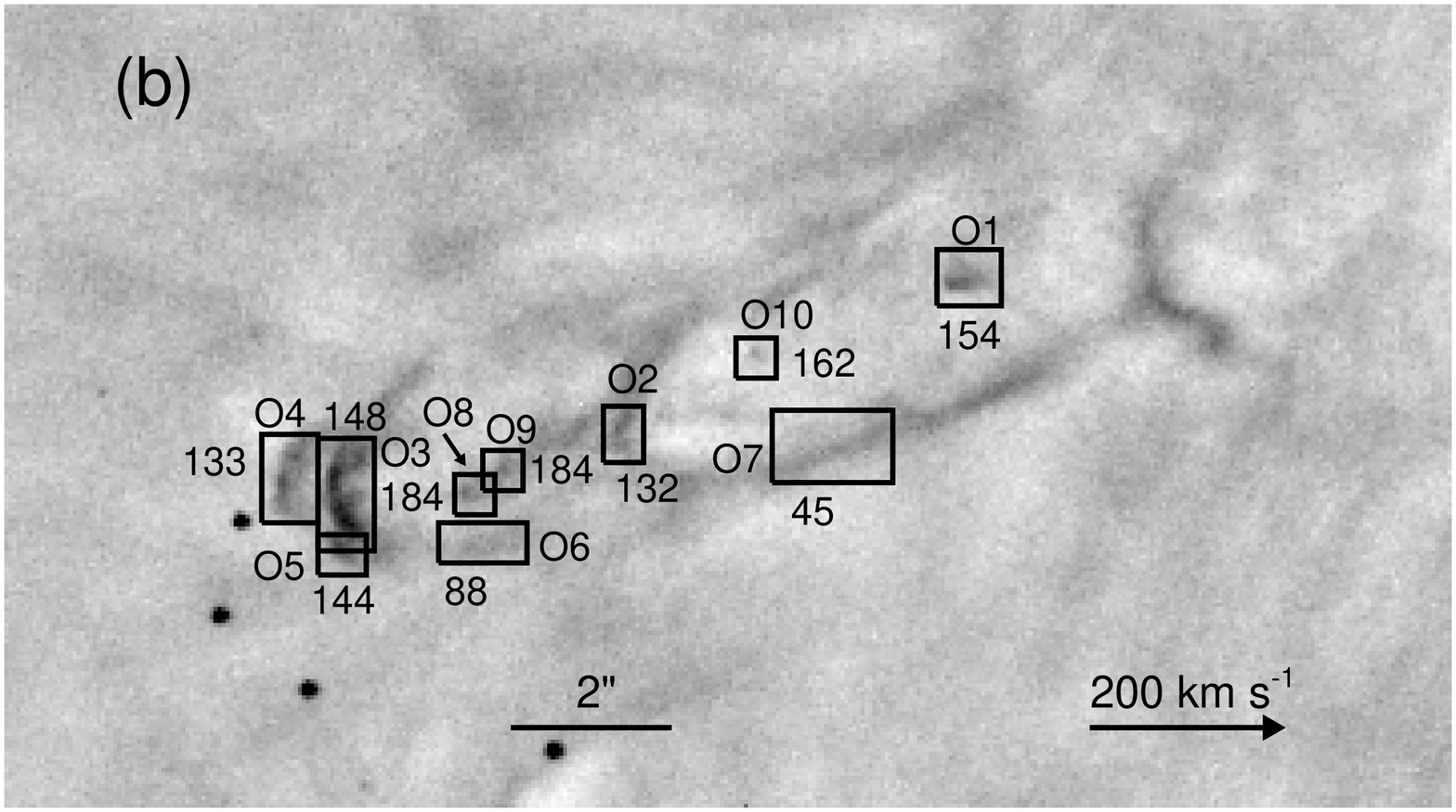}  &
\includegraphics[angle=0,scale=0.35]{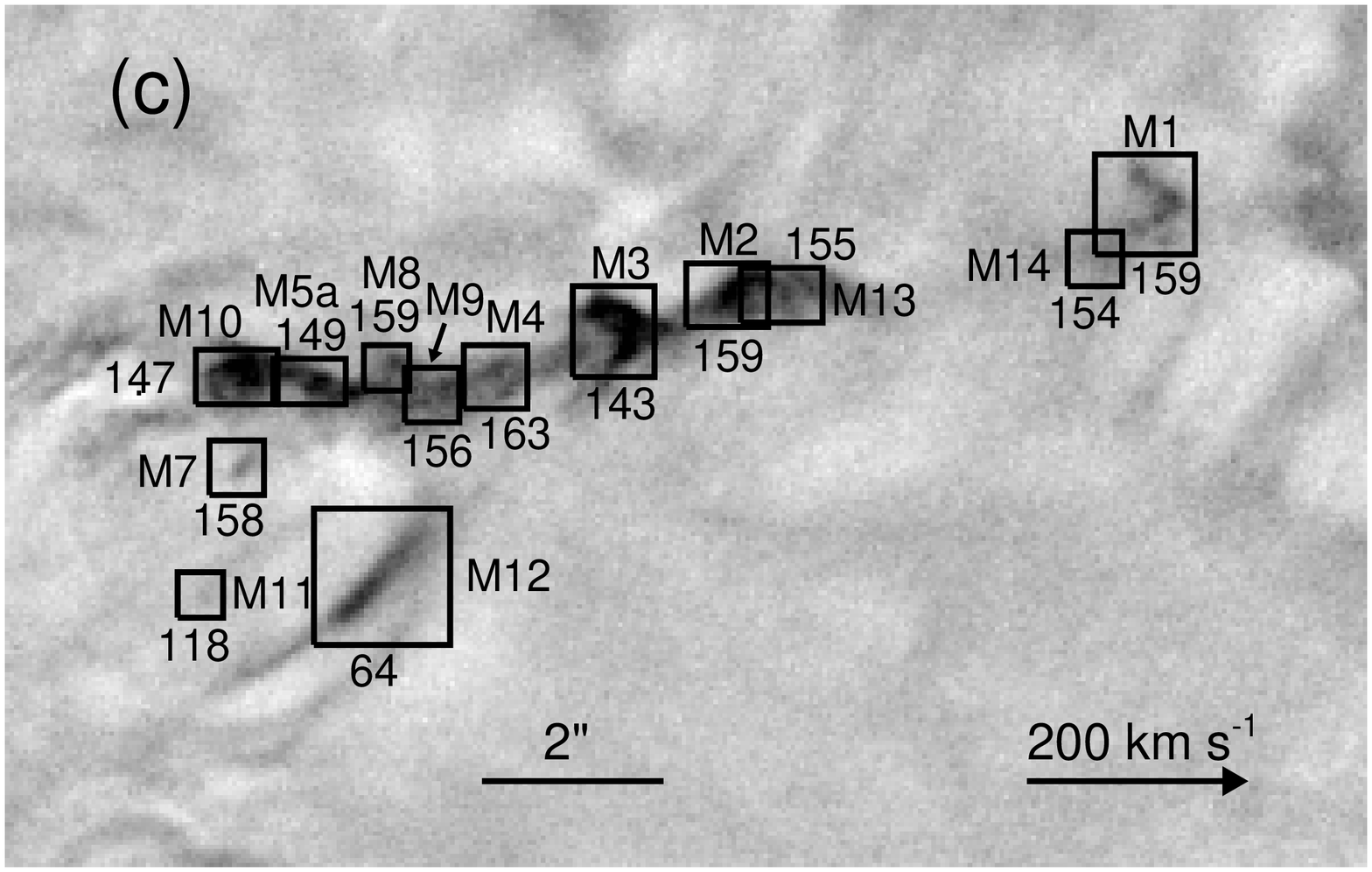}  \\
\includegraphics[angle=0,scale=0.335]{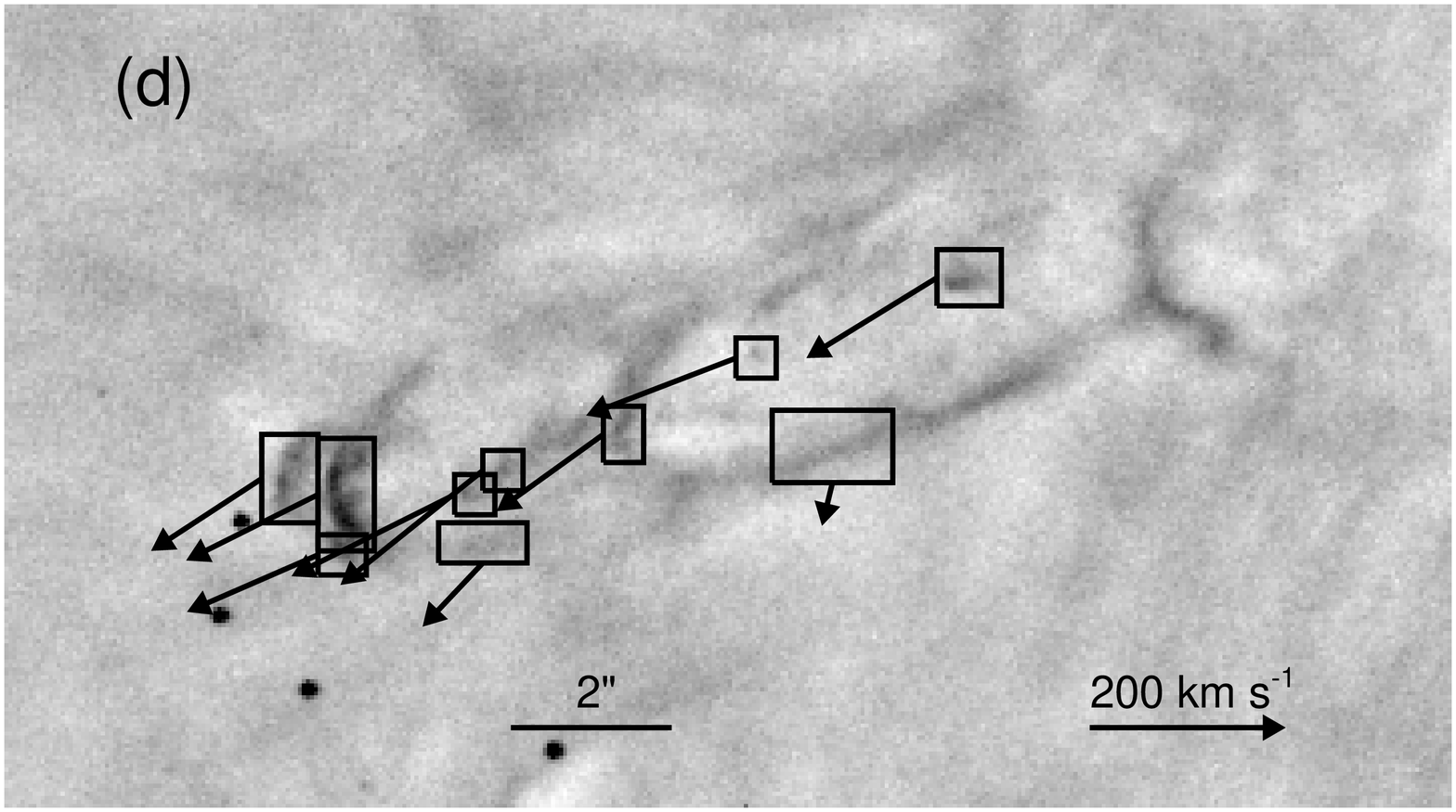}  &
\includegraphics[angle=0,scale=0.35]{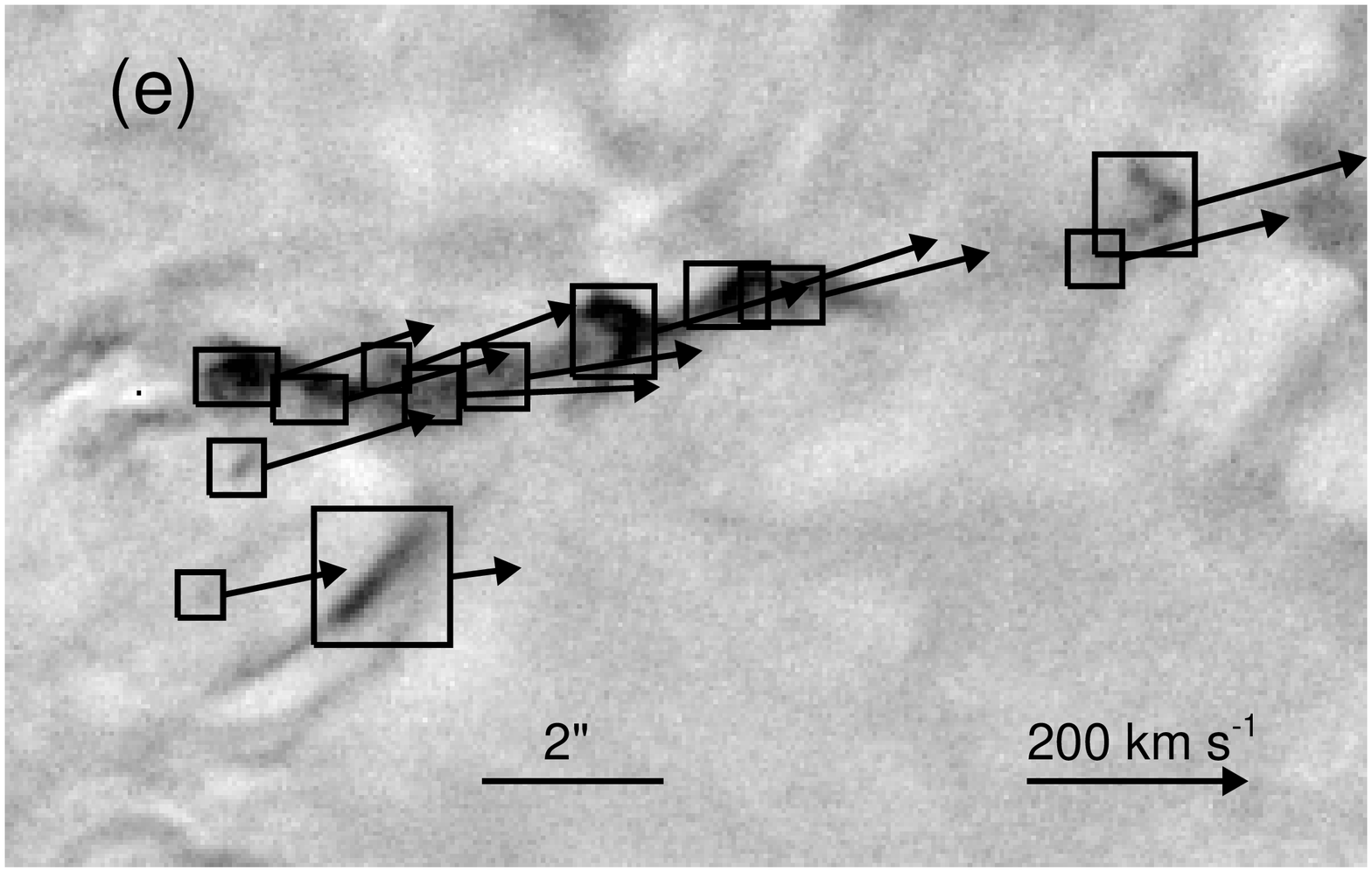}  \\
\end{array}
\end{array}$
\caption{
\textit{Top:} Difference image of the first and second epochs of \textit{HST}/ACS imaging (a). 
\textit{Middle:} 
New \textit{HST}/ACS H$\alpha$ image of HH~666~O (b) and HH~666~M (c). Boxes indicate features used to measure proper motions with the name of the feature and the measured transverse velocity labelled alongside. 
\textit{Bottom:} 
Same boxes as the middle panel with vectors to indicate the magnitude and direction of the measured motion (d and e).
}\label{fig:pm} 
\end{figure*}

With the longer time baseline and better match of the observational setup, we can also measure proper motions of slower features and off-axis motions with greater fidelity.  
The most remarkable motions we uncover with the latest epoch shows clear \textit{lateral} expansion of the cocoon seen in H$\alpha$ emission (see Figure~\ref{fig:pm}). 
In particular, we measure a transverse velocity for feature M12 of 64 km s$^{-1}$ and it appears to move at an angle $\sim 10^{\circ}$ away from the jet axis defined by the motion of features M8, M5a, M10. 
Similar lateral expansion in features O6, O7 from HH~666~O traces the slow expansion of the H$\alpha$ sheath surrounding the [Fe~{\sc ii}] jet. 
A slower velocity and an off-axis propagation direction are predicted for the expanding wings of a bow shock \citep[e.g.][]{ost01}. 

\section{Resolving morphological and kinematic discrepancies}\label{s:discrep}
\begin{figure*}
\centering
$\begin{array}{c}
\includegraphics[angle=0,scale=0.65]{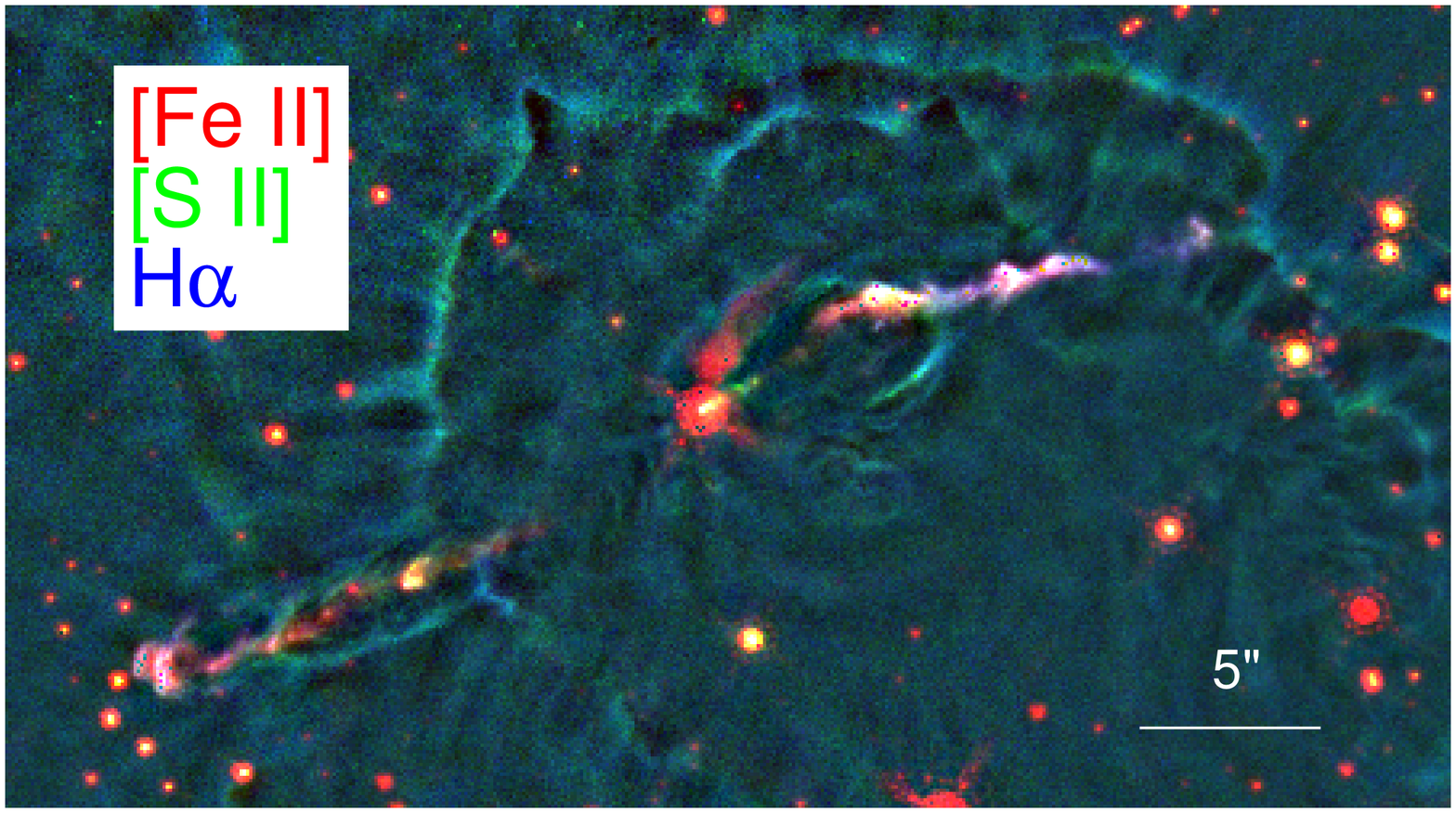} \\
\includegraphics[trim=50mm 50mm 50mm 45mm,angle=90,scale=0.725]{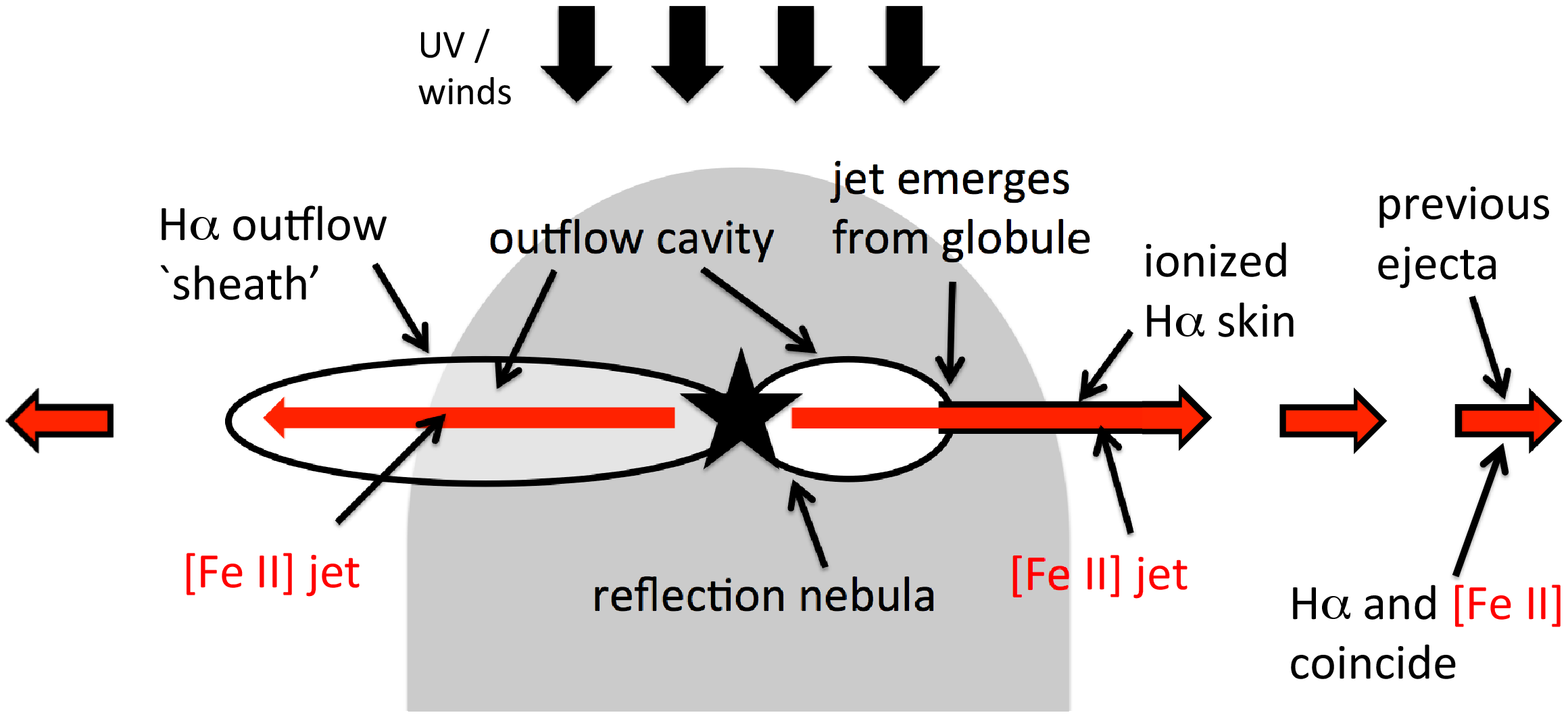} \\ 
\end{array}$
\caption{\textit{Top:} Colour image of HH~666 that illustrates the spatial offset between different emission lines in the jet. 
\textit{Bottom:}
Cartoon of the externally irradiated jet and outflow. Red lines show the collimated [Fe~{\sc ii}] jet body that clears cavities on either side of the jet. H$\alpha$ emission, shown in black, traces the ionized edges of these cavities and the ionized skin of the bare jet. 
}\label{fig:cartoon} 
\end{figure*}
Given the striking disparity in the morphology and kinematics traced by H$\alpha$ and [Fe~{\sc ii}] emission in both images and spectra, we propose that these two different emission lines trace physically distinct components of the HH~666 outflow. 
We illustrate the anatomy of a dense, externally irradiated protostellar outflow in Figure~\ref{fig:cartoon}, and discuss the evidence for this picture below.

\subsection{Redshifted lobe: HH~666~O}\label{ss:o}
In HH~666~O, the redshifted part of the outflow to the southeast of the protostar, 
H$\alpha$ emission appears to form a cocoon that surrounds the [Fe~{\sc ii}] jet core. 
The large spatial offset of the H$\alpha$ emission from the [Fe~{\sc ii}] ($\sim 2$\arcsec) suggests that it traces an entirely different component of the outflow. 
In fact, the thin arcuate morphology of the H$\alpha$ and [S~{\sc ii}] emission along the face of the pillar resembles the molecular outflow lobes seen at longer (mm) wavelengths \citep[e.g. HH~211,][]{gue99,dio10,tap12}. 
Prompt entrainment of ambient material via bow shocks from episodic jets has been invoked to explain nested outflow shells \citep[e.g.][]{nar96}. 
Jets in Carina that are driven by protostars embedded in a dust pillar \citep[more than half the jets discovered by][]{smi10} will entrain material as they exit the cloud. 

Kinematics in the H$\alpha$ spectrum also support the interpretation that H$\alpha$ emission traces an externally irradiated outflow shell in HH~666. 
Fast, ionized material at the head of the bow shock that creates this outflow shell corresponds to the highest velocities seen in the H$\alpha$ spectrum. 
Velocities observed closer to the driving source are lower because they are dominated by emission from the slower wings of the bow shock. 
The slowest velocities will be observed farthest from the apex of the bow shock, and therefore closest to the driving source \citep[see, e.g. Figure~1 in][]{arc02}. 
In this picture, the bow shock dominates entrainment of the ambient gas and will create the observed Hubble-like velocity structure, as seen in the position-velocity diagrams of molecular outflows \citep[e.g. HH~300,][]{arc01}. 
In contrast to the H$\alpha$ emission, Doppler velocities in the collimated [Fe~{\sc ii}] jet are remarkably constant throughout HH~666~O, remaining close to $+200$ km s$^{-1}$, as expected for an unimpeded jet propagating away from its driving source. 
We do not detect evidence for a Hubble-flow in the [Fe~{\sc ii}] emission from the jet. 
If environmental interaction is dominated by the first impulse from the jet, portions of the jet traced by [Fe~{\sc ii}] that travel in the wake of the initial bow shock will transfer little momentum to the environment.

Some optical emission from the jet body in the outer portions of HH~666~O, near the leading bow shock (feature O3 in Figure~\ref{fig:pm}, also see Figure~\ref{fig:hst_ims}) further supports the idea that there are two externally irradiated outflow components visible in HH~666. 
H$\alpha$, [S~{\sc ii}], and [Fe~{\sc ii}] emission begin to coincide just beyond the edge of the dust pillar, marking the first place where the newly liberated \textit{jet} is laid bare to ionizing radiation. 
Optical emission from the \textit{jet} beyond the pillar edge suggests a decrease in the intervening optical depth, and the increase of the [Fe~{\sc ii}] ratio $\mathcal{R}$ toward the terminus of HH~666~O supports this interpretation. 
However, the ratio only approaches the intrinsic value beyond the end of the jet, suggesting that the observed reddening is primarily from the jet itself (see Figure~\ref{fig:feii_ratio}).

\subsection{Blueshifted lobe: HH~666~M}\label{ss:m}
The blueshifted part of the outflow, HH~666~M, that extends to the northwest of the protostar reveals a similar picture to HH~666~O. 
Close to the driving source (within $\lesssim 10$\arcsec), the behavior of the H$\alpha$ and [Fe~{\sc ii}] emission in HH~666~M resembles that of HH~666~O. 
A narrow column of knotty [Fe~{\sc ii}] emission bisects broad, arc-like H$\alpha$ emission before the morphologies traced by the two lines converge at larger distances from the protostar. 
New H$\alpha$ images show that the head of the H$\alpha$ arc moves along the jet axis while its wings expand laterally away from it. 
Features M8, M5a, M10 (see Figure~\ref{fig:pm}) move primarily along the jet axis with a transverse velocity of $\sim 150$ km s$^{-1}$. 
Feature M12 travels at an $\sim 10^{\circ}$ angle away from the jet axis, indicating that this emission feature traces the expansion of the H$\alpha$ arc away from the jet axis. 
Expanding H$\alpha$ emission suggests that the arc may be the bow shock where HH~666~M breaks out of the front side of the parent cloud (approximately at the location of feature M9). 
In this picture, the `dark patch' to the northwest of the driving source is a cavity cleared by the jet as it plows out of the molecular cloud, leaving a deficit of material behind the bow shock. 
\citet{smi10} also suggested that a Cant\'{o} nozzle \citep{can80,can81} may be responsible for refocusing the jet at this point. 

Spectra support the interpretation that the H$\alpha$ arc traces the tip of the expanding outflow cavity. 
We observe a Hubble wedge in HH~666~M, with H$\alpha$ emission tracing increasing velocities up until a sharp drop corresponding to features M5a, M10, M12 (see Figure~\ref{fig:spectra}). 
Inside feature M9, closer to the driving source, there is a clear deficit of emission from the globule. 
However, at the same place along the slit where globule emission decreases, we see blueshifted emission that extends smoothly toward feature M9. 
Thus, it appears that H$\alpha$ emission from the inner portion of HH~666~M also traces the expansion of the outflow shell.

\begin{figure}
\centering
$\begin{array}{c}
\includegraphics[angle=0,scale=0.315]{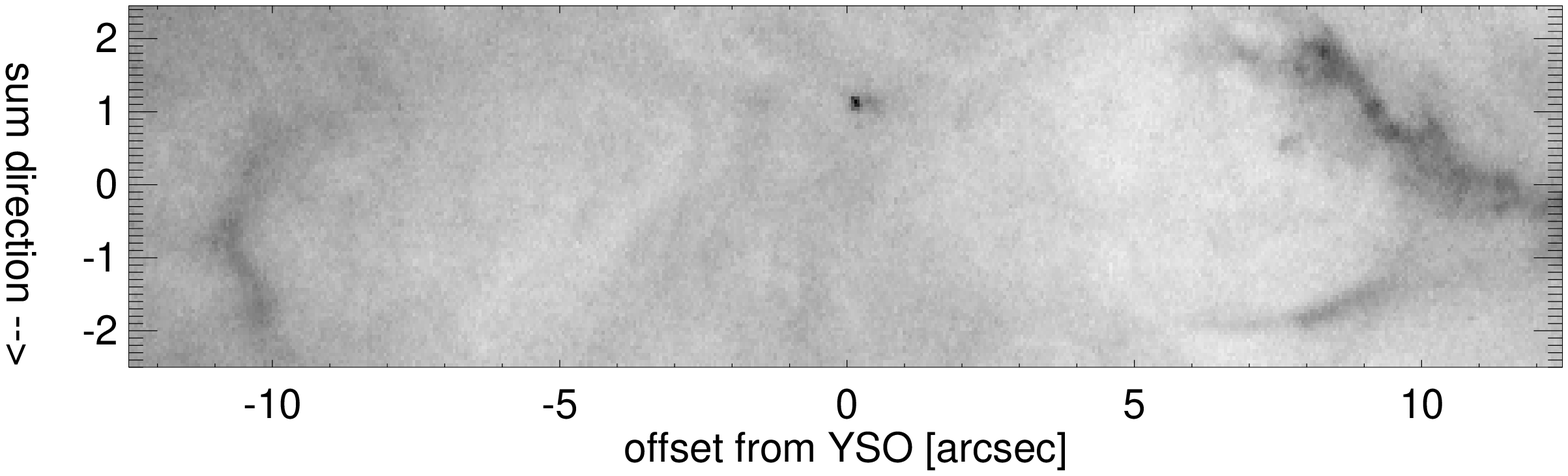} \\ 
\includegraphics[angle=90,scale=0.315]{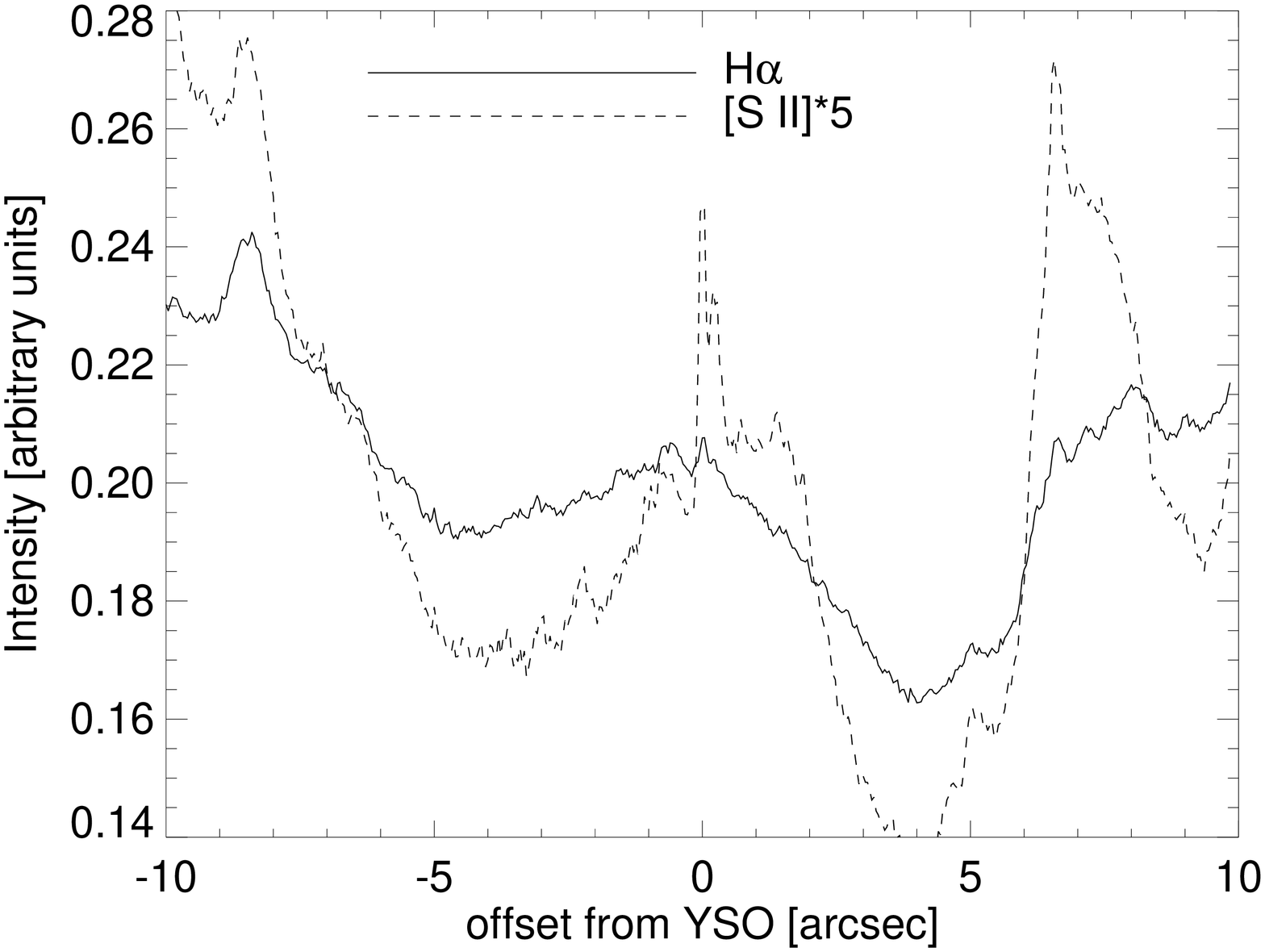} \\ 
\end{array}$
\caption{Tracing through the \textit{HST-WFC3} H$\alpha$ and [S~{\sc ii}] images of HH~666. 
Top panel shows the H$\alpha$ image of the area used to plot the tracings through the jet environment shown below. 
The plotted intensity is the total of the emission through the y-direction of the image above (perpendicular to the jet propagation direction). 
A dark cavity along the northwest limb of the jet is apparent in optical and IR images of the jet. This corresponds to the dip in the flux to the right of the driving source. 
The dip to the left indicates the presence of a complementary evacuated cavity along the southeast limb of the jet (see Section~\ref{ss:m}). 
}\label{fig:cavity_tracings} 
\end{figure}

A Hubble-like velocity pattern in HH~666~M further suggests that the dark patch to the northwest of the driving source is an evacuated cavity in the dust pillar cleared by the jet. 
If this is correct, HH~666~O should create a corresponding cavity to the southeast of the driving source as it propagates deeper into the cloud. 
To test this, we took an intensity tracing through the pillar along the axis of the jet to search for a decrease in the ambient H$\alpha$ emission on either side of the driving source (see Figure~\ref{fig:cavity_tracings}). 
Indeed, intensity tracings reveal dips in the H$\alpha$ intensity along the pillar extending $\sim 5$\arcsec\ on either side of the driving source. 
The emission trough is shallower to the southeast of the driving source, as expected for more foreground emission from the pillar itself.  

As in HH~666~O, Doppler velocities measured in the [Fe~{\sc ii}] spectra of HH~666~M remain steady, with velocities close to $-200$ km s$^{-1}$ throughout. 
Fast, oppositely directed [Fe~{\sc ii}] emission is what we expect from a spatially resolved steady \textit{jet}. 
This structure has been seen in other jets \citep[e.g.][]{gar08,mel09,liu12}, including those driven by intermediate-mass protostars \citep[e.g.][]{ell13,ell14}. 
Hubble-like velocities have been observed in protostellar \textit{outflows}, but usually in IR/mm tracers of molecular gas that has been accelerated by an underlying jet \citep[e.g.][]{arc01}. 
Only two other jets show Hubble-like velocities in H$\alpha$ emission -- HH~83 \citep{rei89} and HH~900 \citep{rei15}.

Beyond the H$\alpha$ arc (west of feature M10, outside the globule), H$\alpha$ and [Fe~{\sc ii}] emission converges, thereafter tracing the same morphology. 
The [Fe~{\sc ii}] ratio remains relatively constant (see Figure~\ref{fig:feii_ratio}), and velocities traced by both lines remain relatively steady (no strong shocks). 
Doppler velocities in [Fe~{\sc ii}] spectra, however, are nearly 100 km s$^{-1}$ faster than those traced by H$\alpha$. 
Low- and high-velocity components have been observed in jets from low-mass T Tauri stars with velocity differences between the core of the jet and the slightly off-axis wings of as much as $\sim 100$ km s$^{-1}$ in some cases \citep[e.g. DG~Tau,][]{bac00,pyo03}. 
If the shell of material that is traced by H$\alpha$ inside the globule is pushed back against the jet by feedback from massive stars in the region, then the two components may overlap spatially in images while the entrained material traced by H$\alpha$ has decelerated as compared to the velocity of the [Fe~{\sc ii}] jet.

\subsection{An externally irradiated jet and outflow}
For embedded jets that entrain an outflow as they plow through a dusty pillar, UV radiation from nearby massive stars will illuminate the shell of the outflow in addition to the jet. 
We argue that H$\alpha$ proper motions and Doppler velocities trace the expansion and propagation of the externally irradiated outflow lobe entrained by the jet, but not the jet core. 
In this picture, we expect larger proper motions from the jet itself ($\sim 18$ mas yr$^{-1}$ for a velocity of 200 km s$^{-1}$), although additional [Fe~{\sc ii}] imaging at a later epoch is required to test this hypothesis.

Our data show that H$\alpha$ emission traces entrained outflowing gas with the same Hubble-like velocity structure seen in molecular outflows associated with other HH jets \citep[e.g.][]{lee01,arc01}. 
However, the extreme UV radiation in the Carina nebula illuminates entrained gas in HH~666, leading to H$\alpha$ emission from the outflow cocoon. 
No H$_2$ emission that is unambiguously associated with the jet has been observed in HH~666 \citep{smi04,har15}, although weak H$_2$ emission will be difficult to discern against the bright background of the photodissociation region (PDR).

\section{An expanding outflow shell from an ongoing jet burst} 

Both morphology and kinematics reveal an H$\alpha$ outflow lobe encasing the [Fe~{\sc ii}] jet to the east and west of the HH~666 driving source. 
Proper motions reveal the dynamical age of these H$\alpha$ outflow lobes to be $\sim 1000-1500$ yrs (see Table~2). 
This points to a sudden increase in the jet mass-loss rate $\sim 1000$ yrs ago that provided a “kick” to the ambient material now seen as a Hubble flow in H$\alpha$ emission. 
Several more distant knots in HH~666 indicate that the jet had other outbursts like this in the past \citep{smi04,smi10}. 

Steady [Fe~{\sc ii}] emission inside the H$\alpha$ cocoons on either side of the jet indicates that the high-mass-loss rate episode has persisted, with a duration comparable to its age. 
A burst duration of $\sim 1000$ yrs is an order of magnitude larger than the estimated decay time of FU~Orionis outbursts \citep[e.g.][]{hk96}. 
This long duration is on the order of the time between ejections found in HH jets driven by low-mass stars \citep[e.g.][]{rei89,har90,rei92}, and similar to the variability time-scale estimated for HH~46/47 \citep{rag92}. 
While the burst duration is long compared to those inferred for low-mass protostars, the $\sim 1000$ yr dynamical time of the inner jet is only a small fraction of the overall age of the HH~666 outflow \citep[$\sim 40,000$ yr, see][]{rei14}. 
This points to a strong, sustained increase in the accretion rate from a relatively evolved intermediate-mass protostar. 
A larger sample of intermediate-mass protostars is required to determine the frequency and duty cycle of such accretion bursts, and whether they remain stronger than their low-mass counterparts throughout their evolution.

[Fe~{\sc ii}] emission from HH~666 traces a steady jet with higher densities and velocities than traced by H$\alpha$ emission. 
\citet{rei13} found that the mass-loss rate implied by the survival of Fe$^+$ in the jet is an order of magnitude higher than the mass-loss rate derived from the H$\alpha$ emission measure \citep{smi10}. 
New [Fe~{\sc ii}] spectra presented here reveal velocities $\pm 200$ km s$^{-1}$ throughout the inner jet, similar to the fastest velocities in the H$\alpha$ spectrum. 
Applying the median tilt angle of the jet derived from the H$\alpha$ velocities, $\alpha \approx 42^{\circ}$ (see Table~2), implies that the three-dimensional velocity of the [Fe~{\sc ii}] jet is $\sim 300$ km s$^{-1}$, almost twice the speed of the H$\alpha$ jet. 
The higher mass-loss rate and velocities in the [Fe~{\sc ii}] jet suggest that this component of the outflow has at least an order of magnitude more momentum. 
However, the steady velocity of the [Fe~{\sc ii}] jet traveling in the wake of the initial bow shock suggests little momentum transfer to the local environment. 
This means that the high momentum of the [Fe~{\sc ii}] jet will not be deposited locally in the molecular globule, but rather injected into the H~{\sc ii} region where its contribution to driving turbulence in the region will be dwarfed by feedback from massive stars.

\section{Conclusions}\label{s:conclusion}

In this paper, we propose a solution to the discrepancy observed in the morphology and kinematics of H$\alpha$ and [Fe~{\sc ii}] emission from HH~666, an externally irradiated protostellar outflow in the Carina nebula. 
Previous high-resolution imaging revealed different morphologies traced by the two lines. 
We present new, near-IR spectra of the inner jet demonstrating that [Fe~{\sc ii}] emission also traces different kinematics. 
Doppler velocities in the new [Fe~{\sc ii}] spectra of the inner jet, features HH~666~M and O, remain remarkably consistent on either side of the jet, tracing velocities of $\pm 200$ km s$^{-1}$ throughout. 
This stands in stark contrast to the velocity structure seen from the same portions of the jet in the longslit H$\alpha$ spectrum from \citet{smi04}. 
Both HH~666~M and O show Hubble-like velocities in the H$\alpha$ spectrum with only the fastest Doppler velocities in the Hubble wedges on either side of the driving source approaching the velocities traced by [Fe~{\sc ii}].

H$\alpha$ and [Fe~{\sc ii}] emission also trace different morphologies, complementing the velocity discrepancies in spectra \citep{rei13}. 
Sheath-like H$\alpha$ emission encompasses narrow [Fe~{\sc ii}] emission on either side of the jet. 
We argue that broad H$\alpha$ emission traces the material swept up via prompt entrainment by a jet outburst that began $\sim 1000$ yrs ago, with bright H$\alpha$ arcs tracing the head of the bow shock that dominates entrainment. 
Constant velocity [Fe~{\sc ii}] emission, inside the outflow sheath and Hubble flow traced by H$\alpha$, indicates that the high-mass-loss rate episode from the jet is ongoing, tracing a strong accretion burst that has been sustained longer than the typical decay time of FU~Orionis outbursts. 
A new epoch of H$\alpha$ imaging from \textit{HST}/ACS provides a $\sim 9$ yr time baseline to measure the motion of bright H$\alpha$ features. 
Proper motions trace the lateral expansion of the H$\alpha$ sheath, consistent with an expanding shell around the jet.

We propose that the same underlying physics that drives the molecular outflows seen from more embedded intermediate-mass protostars creates the two-component outflow we observe in HH~666. 
Unlike jets with molecular outflows, HH~666 resides in a giant H~{\sc ii} region, so both the jet and the material it entrains are subject to the harsh UV background created by the many O-type stars in the Carina nebula. 
This leads to H$\alpha$ emission from the walls of the outflow cavity and [Fe~{\sc ii}] emission from the obscured jet. 
Outside the parent pillar where a cocoon no longer envelopes the jet, H$\alpha$ and [Fe~{\sc ii}] emission trace the same morphology. 
Thus, HH~666 provides a missing link that unifies the irradiated outflows driven by unobscured protostars with the molecular outflows typically observed from embedded intermediate-mass protostars.


\section*{Acknowledgments}
We would like to thank Rob Simcoe for his assistance with reduction of the FIRE data. Support for this work was provided by NASA grant AR-12155, GO-13390, and GO-13391 from the Space Telescope Science Institute. 
This work is based on observations made with the NASA/ESA Hubble Space Telescope, obtained from the Data Archive at the Space Telescope Science Institute, which is operated by the Association of Universities for Research in Astronomy, Inc., under NASA contract NAS 5-26555. These \textit{HST} observations are associated with programs GO~10241, 10475, 13390, and 13391. 
Gemini observations are from the GS-2013A-Q-12 science program.



\label{lastpage}


\begin{thebibliography}{}
\bibitem[\protect\citeauthoryear{Anderson}{2006}]{and05} Anderson J., 2006, ``The 2005 HST calibration workshop: Hubble after the transition to two-gyro mode'', Proceedings of a workshop held at the Space Telescope Science Institute, Baltimore, Maryland, October 26-28, 2005. Eds Koekemoer A.M., Goudfrooij P., Dressel L.L. Greenbelt, MD, 11
\bibitem[\protect\citeauthoryear{Anderson \& King}{2006}]{and06} Anderson J., King I.R., 2006, ``PSFs, Photometry, and Astronomy for the ACS/WFC'', 2006, 1
\bibitem[\protect\citeauthoryear{Anderson et al.}{2008a}]{and08a} Anderson J., et al., 2008a, AJ, 135, 2055
\bibitem[\protect\citeauthoryear{Anderson et al.}{2008b}]{and08b} Anderson J., et al., 2008b, AJ, 135,2114
\bibitem[\protect\citeauthoryear{Anderson \& van der Marel}{2010}]{and10} Anderson J., van der Marel R.P., 2010, ApJ, 710,1032
\bibitem[\protect\citeauthoryear{Arce \& Goodman}{2001a}]{arc01a} Arce H.G., Goodman A.A., 2001, ApJ, 554, 132
\bibitem[\protect\citeauthoryear{Arce \& Goodman}{2001b}]{arc01} Arce H.G., Goodman A.A., 2001, ApJL, 551, 171
\bibitem[\protect\citeauthoryear{Arce \& Goodman}{2002}]{arc02} Arce H.G., Goodman A.A., 2002, ApJ, 575, 928
\bibitem[\protect\citeauthoryear{Arce \& Sargent}{2005}]{arc05} Arce H.G., Sargent A.I., 2005, ApJ, 624, 232
\bibitem[\protect\citeauthoryear{Arce \& Sargent}{2006}]{arc06} Arce H.G., Sargent A.I., 2006, ApJ, 646, 1070 
\bibitem[\protect\citeauthoryear{Arce et al.}{2013}]{arc13} Arce H.G., Mardones D., Corder S.A., Garay G., Noriega-Crespo A., Raga A.C., 2013, ApJ, 774, 39
\bibitem[\protect\citeauthoryear{Bacciotti et al.}{2000}]{bac00} Bacciotti F., Mundt R., Ray T.P., Eisl\"{o}ffel J., Solf J., Camezind M., 2000, ApJ, 537, 49
\bibitem[\protect\citeauthoryear{Bally \& Reipurth}{2001}]{bal01} Bally J., Reipurth B., 2001, ApJ, 546, 299 
\bibitem[\protect\citeauthoryear{Bally et al.}{2006}]{bal06} Bally J., Licht D., Smith N., Walawender J., 2006, AJ, 131, 473
\bibitem[\protect\citeauthoryear{Beltr\'{a}n et al.}{2008}]{bel08} Beltr\'{a}n M.T., Estalella R., Girart J.M., Ho P.T.P., Anglada G., 2008, A\&A, 481, 93  
\bibitem[\protect\citeauthoryear{Beuther et al.}{2008}]{beu08} Beuther H., Walsh A.J., Thorwirth S., Zhang Q., Hunter T.R., Megeath S.T., Menten K.M., 2008, A\&A, 481, 169 
\bibitem[\protect\citeauthoryear{Cant\'{o} \& Rodriguez}{1980}]{can80} Cant\'{o} J., Rodriguez L.F., 1980, ApJ, 239, 982
\bibitem[\protect\citeauthoryear{Cant\'{o} et al.}{1981}]{can81} Cant\'{o} J., Rodriguez L.F., Barral J.F., Carral P., 1981, ApJ, 244, 102
\bibitem[\protect\citeauthoryear{Caratti o Garatti et al.}{2008}]{car08} Caratti o Garatti A., Froebrich D., Eisl\"{o}ffel J., Giannini T., Nisini B., 2008, A\&A, 485, 137 
\bibitem[\protect\citeauthoryear{Currie et al.}{1996}]{cur96} Currie D.G. et al., 1996, AJ, 112, 1115 
\bibitem[\protect\citeauthoryear{De Gouveia et al.}{1996}]{deG96} De Gouveia E.M., Birkinshaw M., Benz W., 1996, ApJL, 460, 111
\bibitem[\protect\citeauthoryear{Dionatos et al.}{2010}]{dio10} Dionatos O., Nisini B., Cabrit S., Kristensen L., Pineau des For\^{e}ts G., 2010, A\&A, 521, 7
\bibitem[\protect\citeauthoryear{Duarte-Cabral et al.}{2013}]{dua13} Duarte-Cabral A., Bontemps S., Motte F., Hennemann M., Schneider N., Andr\'{e} P., 2013, A\&A, 558, 125 
\bibitem[\protect\citeauthoryear{Ellerbroek et al.}{2013}]{ell13} Ellerbroek L.E., Podio L., Kaper L., Sana H., Huppenkothen D., de Koter A., Monaco L., 2013, A\&A, 551, 5
\bibitem[\protect\citeauthoryear{Ellerbroek et al.}{2014}]{ell14} Ellerbroek L.E., et al., 2014, A\&A, 563, 87
\bibitem[\protect\citeauthoryear{Forbrich et al.}{2009}]{for09} Forbrich J., Stanke Th., Klein R., Henning Th., Menten K.M., Schreyer K., Posselt B., 2009, A\&A, 493, 547
\bibitem[\protect\citeauthoryear{Fuente et al.}{2009}]{fue09} Fuente A. et al., 2009, A\&A, 507, 1475
\bibitem[\protect\citeauthoryear{Garay et al.}{2003}]{gar03} Garay G., Brooks K.J., Mardones D., Norris R.P., 2003, ApJ, 587, 739 
\bibitem[\protect\citeauthoryear{Garcia-Lopez et al.}{2008}]{gar08} Garcia-Lopez R., Nisini B., Giannini T., Eisl\"{o}ffel J., Bacciotti F., Podio L., 2008, A\&A, 487, 1019 
\bibitem[\protect\citeauthoryear{Giannini et al.}{2015}]{gia15} Giannini, T., et al., 2015, ApJ, 798, 33 
\bibitem[\protect\citeauthoryear{Gueth \& Guilloteau}{1999}]{gue99} Gueth F., Guilloteau S., 1999, A\&A, 343, 571
\bibitem[\protect\citeauthoryear{Guzm\'{a}n et al.}{2012}]{guz12} Guzm\'{a}n A.E., Garay G., Brooks K.J., Voronkov M.A., 2012, ApJ, 753, 51
\bibitem[\protect\citeauthoryear{Hartigan et al.}{1990}]{har90} Hartigan P., Raymond J., Meaburn J., 1990, ApJ, 362, 624  
\bibitem[\protect\citeauthoryear{Hartigan et al.}{1994}]{har94} Hartigan P., Morse J.A., Raymond J., 1994, ApJ, 436, 125 
\bibitem[\protect\citeauthoryear{Hartigan et al.}{2001}]{har01} Hartigan P., Morse J.A., Reipurth B., Heathcote S., Bally J., 2001, ApJL, 559, 157
\bibitem[\protect\citeauthoryear{Hartigan et al.}{2011}]{har11} Hartigan P., et al., 2011, ApJ, 736, 29
\bibitem[\protect\citeauthoryear{Hartigan et al.}{2015}]{har15} Hartigan P., Reiter M., Smith N., Bally J., 2015, AJ, 149, 101
\bibitem[\protect\citeauthoryear{Hartmann \& Kenyon}{1996}]{hk96} Hartmann L., Kenyon, S.J., 1996, ARA\&A, 34, 207 
\bibitem[\protect\citeauthoryear{Heathcote et al.}{1996}]{hea96} Heathcote S., Morse J.A., Hartigan P., Reipurth B., Schwartz R.D., Bally J., Stone J.M., 1996, AJ, 112, 1141
\bibitem[\protect\citeauthoryear{Klaassen et al.}{2013}]{kla13} Klaassen P.D., et al. 2013, A\&A, 555, 73 
\bibitem[\protect\citeauthoryear{Lee et al.}{2001}]{lee01} Lee C.-F., Stone J.M., Ostriker E.C., Mundy L.G., 2001, ApJ, 557, 429
\bibitem[\protect\citeauthoryear{LeFloch et al.}{2007}]{lef07} Lefloch B., Cernicharo J., Reipurth B., Pardo J.R., Neri R., 2007, ApJ, 658, 498
\bibitem[\protect\citeauthoryear{Liu et al.}{2012}]{liu12} Liu C.-F., et al. 2012, ApJ, 749, 62
\bibitem[\protect\citeauthoryear{McGroarty et al.}{2004}]{mcg04} McGroarty F., Ray T.P., Bally J., 2004, A\&A, 415, 189
\bibitem[\protect\citeauthoryear{Melnikov et al.}{2009}]{mel09} Melnikov S.Y., Eisl\"{o}ffel J., Bacciotti F., Woitas J., Ray T.P., 2009, A\&A, 506, 763 
\bibitem[\protect\citeauthoryear{Morse et al.}{2001}]{mor01} Morse J.A., Kellogg J.R., Bally J., Davidson K., Balick B., Ebbets D., 2001, ApJL, 548, 207 
\bibitem[\protect\citeauthoryear{Mundt et al.}{1991}]{mun91} Mundt R., Ray T.P., Raga A.C., 1991, A\&A, 252, 740 
\bibitem[\protect\citeauthoryear{Nagar et al.}{1997}]{nag97} Nagar N.M., Vogel S.N., Stone J.M., Ostriker E.C., 1997, ApJ, 482, 195
\bibitem[\protect\citeauthoryear{Narayanan \& Walker}{1996}]{nar96} Narayanan G., Walker C.K., 1996, ApJ, 466, 844 
\bibitem[\protect\citeauthoryear{Noriega-Crespo et al.}{2004}]{nor04} Noriega-Crespo A., et al., 2004, ApJS, 154,352
\bibitem[\protect\citeauthoryear{Ostriker et al.}{2001}]{ost01} Ostriker E.C., Lee C.-F., Stone J.M., Mundy L.G., 2001, ApJ, 557, 443 
\bibitem[\protect\citeauthoryear{Povich et al.}{2011}]{pov11} Povich M. et al., 2011, ApJS, 194, 14 
\bibitem[\protect\citeauthoryear{Pyo et al.}{2003}]{pyo03} Pyo T.-S., et al., 2003, ApJ, 590, 340
\bibitem[\protect\citeauthoryear{Raga \& Noriega-Crespo}{1992}]{rag92} Raga A.C., Noriega-Crespo A., 1992, Rev. Mex. Astron. Astrofis., 24, 9 
\bibitem[\protect\citeauthoryear{Raga \& Cabrit}{1993}]{rag93c} Raga A.C., Cabrit S., 1993, A\&A, 278, 267  
\bibitem[\protect\citeauthoryear{Reipurth et al.}{1989}]{rei89} Reipurth B., 1989, A\&A, 220, 249 
\bibitem[\protect\citeauthoryear{Reipurth et al.}{1992}]{rei92} Reipurth B., Raga A.C., Heathcote S., 1992, ApJ, 392, 145
\bibitem[\protect\citeauthoryear{Reipurth et al.}{1997a}]{rei97} Reipurth B., Hartigan P., Heathcote S., Morse J.A., Bally J., 1997, AJ, 114, 757 
\bibitem[\protect\citeauthoryear{Reipurth et al.}{1998}]{rei98} Reipurth B., Bally J., Fesen R.A., Devine D., 1998, Nature, 396, 343
\bibitem[\protect\citeauthoryear{Reipurth et al.}{2000}]{rei00} Reipurth B., Yu K.C., Heathcote S., Bally J., Rodr\'{i}guez L.F., 2000, AJ, 120, 1449 
\bibitem[\protect\citeauthoryear{Reiter \& Smith}{2013}]{rei13} Reiter M., Smith, N., 2013, MNRAS, 433, 2226
\bibitem[\protect\citeauthoryear{Reiter \& Smith}{2014}]{rei14} Reiter M., Smith N., 2014, MNRAS, 445, 3939 
\bibitem[\protect\citeauthoryear{Reiter et al.}{2015}]{rei15} Reiter M., Smith N., Kiminki M.M., Bally J., Anderson J. 2015, MNRAS, accepted
\bibitem[\protect\citeauthoryear{Salyk et al.}{2014}]{sal14} Salyk C., Pontoppidan K., Corder S., Mu\~{n}oz D., Zhang K., Blake G.A., 2014, ApJ, 792, 68
\bibitem[\protect\citeauthoryear{Sep\'{u}lveda et al.}{2011}]{sep11} Sep\'{u}lveda I., Anglada G., Estalella R., L\'{o}pez R., Girart J.M., Yang J., 2011, A\&A, 527, 41 
\bibitem[\protect\citeauthoryear{Shang et al.}{2006}]{sha06} Shang H., Allen A., Li Z.-Y., Liu C.-F., Chou M.-Y., Anderson J., 2006, ApJ, 649, 845 
\bibitem[\protect\citeauthoryear{Simcoe et al.}{2013}]{sim13} Simcoe R.A. et al., 2013, PASP, 125, 270
\bibitem[\protect\citeauthoryear{Smith et al.}{2004}]{smi04} Smith N., Bally J., Brooks K.J., 2004, AJ, 127, 2793 
\bibitem[\protect\citeauthoryear{Smith}{2006a}]{smi06a} Smith N., 2006a, MNRAS, 367, 763
\bibitem[\protect\citeauthoryear{Smith}{2006b}]{smi06} Smith N., 2006b, ApJ, 644, 1151
\bibitem[\protect\citeauthoryear{Smith \& Hartigan}{2006}]{sh06} Smith N., Hartigan P., 2006, ApJ, 638, 1045
\bibitem[\protect\citeauthoryear{Smith et al.}{2010}]{smi10} Smith N., Bally J. Walborn N., 2010, MNRAS, 405, 1153 
\bibitem[\protect\citeauthoryear{Sohn et al.}{2012}]{soh12} Sohn S.T., Anderson J., van der Marel R.P., 2012, ApJ, 753, 7
\bibitem[\protect\citeauthoryear{Takahaski \& Ho}{2012}]{tak12} Takahashi S., Ho P.T.P., 2012, ApJ, 745, 10
\bibitem[\protect\citeauthoryear{Tappe et al.}{2012}]{tap12} Tappe A., Forbrich J., Mart\'{i}n S., Yuan Y., Lada C.J., 2012, ApJ, 751, 9
\bibitem[\protect\citeauthoryear{Zapata et al.}{2014}]{zap14} Zapata L.A., Arce H.G., Brassfield E., Palau A., Patel N., Pineda J.E., 2014, MNRAS, 441, 3696
\bibitem[\protect\citeauthoryear{Zinnecker et al.}{1998}]{zin98} Zinnecker H., McCaughrean M.J., Rayner J.T., 1998, Nature, 394, 862
\end{thebibliography}
\end{document}